\documentclass[acmsmall]{acmart}

\AtBeginDocument{
  \providecommand\BibTeX{{
    \normalfont B\kern-0.5em{\scshape i\kern-0.25em b}\kern-0.8em\TeX}}}

\setcopyright{acmlicensed}
\acmJournal{TOSEM}
\acmYear{2023}

\usepackage{multirow}
\usepackage{arydshln}
\usepackage{subcaption}
\usepackage{colortbl}
\usepackage{amsfonts}

\newcommand{\gcy}[1]{\textcolor{black} {#1}}
 
\newcommand{\postnum}{29,057} 
\newcommand{\samplednum}{2,364}
\newcommand{\usernum}{722,389} 

\newcommand{\finding}[2]{
    \begin{center}
    \fcolorbox{black}{gray!10}{\parbox{.97\linewidth}{
    {#2}
    }}
    \end{center}
}

\begin{document}

\title{An Empirical Study on Challenges for LLM Application Developers}

\author{Xiang Chen}
\authornote{Corresponding Authors.}
    \email{xchencs@ntu.edu.cn}
    \orcid{0000-0002-1180-3891}
\author{Chaoyang Gao}
    \email{gcyol@outlook.com}
    \orcid{0009-0006-2627-1513}
\affiliation{
    \institution{School of Artificial Intelligence and Computer Science, Nantong University}
    \city{Nantong}
  \country{China}
}

\author{Chunyang Chen}
\email{chun-yang.chen@tum.de}
\orcid{0000-0003-2011-9618}
\affiliation{
\institution{Department of Computer Science, Technical University of Munich}
\city{Heilbronn}
  \country{Germany}
}

\author{Guangbei Zhang}
    \email{guangbei0324@gmail.com}
    \orcid{0009-0009-9296-9786}
\affiliation{
    \institution{School of Artificial Intelligence and Computer Science, Nantong University}
    \city{Nantong}
  \country{China}
}

\author{Yong Liu}
\authornotemark[1]
\affiliation{
  \institution{College of Information Science and Technology, Beijing University of Chemical Technology}
  \city{Beijing}
  \country{China}}
\email{lyong@mail.buct.edu.cn}
\orcid{0000-0003-1754-3039}

\renewcommand{\shortauthors}{Chen, et al.}

\begin{abstract}
In recent years, large language models (LLMs) have seen rapid advancements, significantly impacting various fields such as computer vision, natural language processing, and software engineering.
These LLMs, exemplified by OpenAI's ChatGPT, have revolutionized the way we approach language understanding and generation tasks. However, in contrast to traditional software development practices, LLM development introduces new challenges for AI developers in design, implementation, and deployment. These challenges span different areas (such as prompts, APIs, and plugins), requiring developers to navigate unique methodologies and considerations specific to LLM application development.

Despite the profound influence of LLMs, to the best of our knowledge, these challenges have not been thoroughly investigated in previous empirical studies. To fill this gap, we present the first comprehensive study on understanding the challenges faced by LLM developers. Specifically, we crawl and analyze {\postnum} relevant questions from a popular OpenAI developer forum. We first examine their popularity and difficulty. After manually analyzing {\samplednum} sampled questions, we construct a taxonomy of challenges faced by LLM developers.
Based on this taxonomy, we summarize a set of findings and actionable implications for LLM-related stakeholders, including developers and providers (especially the OpenAI organization).

\end{abstract}

\ccsdesc[500]{Software and its engineering~Software creation and management}
\ccsdesc[500]{Computing methodologies~Artificial intelligence}

\keywords{Mining Software Repository, Empirical Study, LLM Developer, Development Challenges, Prompt Engineering}

\maketitle

\section{Introduction}
\label{sec:intro}

Emerging large language models (LLMs) are increasingly attracting attention and becoming hot research topics in computer science. 
Until now, LLMs have demonstrated promising performance in different fields, such as computer vision, natural language processing, and software engineering~\cite{zhao2023survey,wang2024software,fan2023large}.
With the rise in popularity of ChatGPT developed by the OpenAI organization and the rapid development of related LLM technologies~\cite{aydin2023chatgpt,nazir2023comprehensive}, an increasing number of developers are utilizing these techniques to assist in their development processes. 
However, during LLM development, developers often encounter various challenges. For example, correctly configuring and invoking LLM's API can be difficult, including setting parameters, managing rate limits, and handling errors. Developing plugins and applications can be daunting for those unfamiliar with AI and LLMs, involving integration, performance optimization, and ensuring security. Ensuring data privacy and security while handling user data is crucial. Developers must comply with relevant regulations and implement necessary security measures.
Therefore, compared to traditional software development, LLM developers face the following unique challenges:
(1) LLMs can automatically handle many tasks, such as text generation, image recognition, and speech recognition. In contrast, traditional software engineering often requires manual coding of various algorithms and logic to accomplish these tasks.
(2) LLMs can produce variable and sometimes unpredictable outputs, unlike traditional software which typically has deterministic outcomes. Developers must account for this uncertainty and implement mechanisms to manage it.
(3) LLM development often involves working with large-scale datasets to train or fine-tune models. This requires specialized knowledge in data preprocessing, management, and the efficient use of computational resources.
(4) Using LLMs requires a significant amount of data for training and fine-tuning, raising concerns about the privacy and security of user data. Developers need to design effective methods to ensure data security and privacy.
(5) Optimizing the performance of LLMs, such as output accuracy, differs from optimizing traditional software.
(6) Understanding and interpreting the outputs of LLMs can be complex. Developers need to ensure that the model's outputs are reliable and contextually appropriate.
These challenges require a different set of skills and considerations, highlighting the distinct nature of LLM application development compared to traditional software development. However, to the best of our knowledge, these challenges faced by LLM application developers have not been thoroughly investigated in previous empirical studies.

To fill this gap, we present the first comprehensive study on understanding the challenges for LLM developers.
Our empirical study aims to help developers to avoid common pitfalls and improve development efficiency.
To this end, we crawl and analyze {\postnum} questions from the OpenAI developer forum\footnote{\url{https://community.openai.com}}.
The OpenAI developer forum is a collaborative space for LLM developers to seek assistance and share insights on using OpenAI technologies. It supports developers of all skill levels and offers discussions on API integration, plugin development, and best practices. The forum also provides updates on the latest advancements and tools from OpenAI. 

In our empirical study, we want to answer the following four research questions.

\textbf{RQ1: What is the popularity trend of LLM development among developers?}

\textbf{Result.} After analyzing posts related to LLM development, we examine the number of new posts and new users added over each period. Our findings indicate that LLM development is attracting increasing attention from developers, particularly with the introduction of OpenAI pivotal products like ChatGPT. This growing trend underscores the rising popularity of LLM technologies, highlighting the timeliness and importance of our empirical study.

\textbf{RQ2: How difficult is LLM development for developers?}

\textbf{Result.} 

To comprehensively illustrate the difficulty level of the challenges faced by LLM developers, we analyze the number of replies to LLM development-related questions, the proportion of posts with an accepted answer, and the average time required to receive the first reply. The results show that 54\% of the questions receive fewer than three replies, only 8.98\% of posts have an accepted answer, and it takes an average of at least 147 hours (about six days) to receive the first reply. These findings indicate that questions faced by LLM developers are often difficult to resolve.

\textbf{RQ3: What specific challenges for LLM developers?}

\textbf{Result.} We perform a manual analysis on {\samplednum} sampled questions and construct a taxonomy of challenges consisting of 26 categories. 
For example, \emph{Prompt Design}, \emph{Integration with Custom Applications}, and \emph{Token Limitation}. In addition, based on this taxonomy, we summarize findings and actionable implications for LLM stakeholders (such as developers and providers).
By comparing our findings to previous empirical studies on ML (Machine Learning)/DL (Deep Learning) developers~\cite{alshangiti2019developing,morovati2024common,zhang2019empirical}, we reveal that LLM developers encounter unique challenges. For example, prompt design plays a critical role, as developers must carefully craft and refine prompts to optimize LLM output, a process that differs greatly from traditional ML/DL’s focus on feature engineering. Additionally, LLMs often exhibit hallucinations, generating incorrect or non-existent information, which is a less frequent issue in ML/DL models. Another significant challenge is the reproducibility of results, where uncertainty in the generation process or LLM updates can lead to inconsistent outputs. Lastly, the high cost of API calls adds a significant financial burden, a challenge not commonly encountered in traditional ML/DL workflows.

\textbf{RQ4: Can our methodology analyze LLM developer challenges on different platforms?}

\textbf{Result.} 
Though LLM developer challenges are primarily analyzed by mining the OpenAI developer forum, our study has significant implications beyond the scope of OpenAI's LLMs. To verify this, we reconstruct the challenge taxonomy by analyzing issues related to more LLMs (such as Llama provided by Meta AI, and Gemini provided by Google) on GitHub. Therefore, our study's findings can broadly apply to LLM products developed by other companies. The final results show that the commonalities in the challenges developers face (such as prompt design, the hallucination problem, and result reproducibility) are not unique to OpenAI. Still, they are prevalent across the entire LLM ecosystem.

In summary, the insights gained from our research can inform best practices, guide the development of support tools, and improve documentation and community resources for developers working with any LLM-based applications.

To the best of our knowledge, the main contributions of our empirical study can be summarized as follows:

\begin{itemize}
\item We conduct the first empirical study to investigate the challenges faced by LLM application developers by mining the OpenAI developer forum.
\item Our study constructs a taxonomy with detailed findings on these challenges.
\item We show the generalization of our methodology in analyzing LLM developers' challenges by mining other platforms, such as GitHub. 
\item We share our collected posts, analysis scripts, and taxonomy of challenges in a GitHub repository for open science\footnote{\url{https://github.com/judeomg/OADF}}.
\end{itemize}
    
\textbf{Paper Organization.}
The rest of this paper is organized as follows.
Section~\ref{sec:background} introduces the background of LLM development.
Section~\ref{sec:methodology} describes our methodology for collecting and analyzing forum posts.
Section~\ref{sec:popularity} analyzes the popularity of the forum.
Section~\ref{sec:difficult} analyzes the difficulty levels of the posts in the forum.
Section~\ref{sec:taxonomy} details our constructed taxonomy and discusses related findings.
Section~\ref{sec:github} discusses the generalization of our methodology by analyzing issues from more LLMs on GitHub.
Section~\ref{sec:threat} first reviews and synthesizes the key findings and offers actionable implications. Then we analyze potential threats and the efforts to alleviate these threats.
Section~\ref{sec:related} discusses work related to our study.
Section~\ref{sec:conclusion} summarizes our empirical study.

\section{Background}
\label{sec:background}

The development of large language models has revolutionized the field of computer vision, natural language processing, and even software engineering~\cite{zhao2023survey,wang2024software,fan2023large}.
As a leading organization in LLM research, OpenAI offers a range of APIs that enable developers to integrate advanced LLM capabilities into their applications. The OpenAI API provides access to various LLMs that can perform a wide array of tasks, from natural language understanding and generation to more complex tasks like code generation/completion and image recognition. This section provides an overview of the API usage and the development of plugins within the OpenAI ecosystem.

\textbf{API Service.} The API service of the LLM allows developers to leverage the power of state-of-the-art LLMs through a simple interface. Key features of the API include:
(1) Developers can integrate the API into their applications with minimal setup, using familiar HTTP requests.
(2) The API supports various tasks including text generation, translation, summarization, and more.
(3) The API is designed to handle a wide range of requests, from small-scale personal projects to large-scale enterprise applications.
The API offers several endpoints corresponding to different models and capabilities. 
For instance, the davinci model is known for its high performance in generating coherent and contextually relevant text, making it suitable for applications requiring sophisticated language understanding.

\textbf{Plugin Development.} Plugins are extensions that enhance the functionality of software applications by embedding AI capabilities directly within them. For example, OpenAI provides robust support for plugin development, allowing developers to create custom plugins that interact with models provided by OpenAI.
Key aspects of plugin development include: (1) 
Plugins typically make use of the OpenAI API to fetch responses from AI models based on user input or other triggers.
(2) Developers can tailor the behavior of plugins to meet specific needs, such as automating customer support responses, generating content, or providing real-time data analysis.
(3) OpenAI plugins can be integrated with various platforms and services, enhancing the AI capabilities of a wide range of applications from web and mobile apps to enterprise systems.

In November 2023, OpenAI introduced a customized version of ChatGPT called GPTs\footnote{\url{https://openai.com/index/introducing-gpts}} and developed GPT Store\footnote{\url{https://openai.com/index/introducing-the-gpt-store/}} for sharing developed GPTs. Similar to plugins, GPTs serve personalized user needs. Developers can create different GPTs for specific purposes, which can be shared with others. GPTs offer a new way to create a tailored version of ChatGPT, assisting in daily life and specific tasks. For instance, GPTs can assist in analyzing math problems, learning soccer rules, or even helping in designing icons. Moreover, creating GPTs is a simple process that does not require coding, allowing anyone to easily build their GPTs by engaging in conversation with ChatGPT, providing instructions and additional knowledge, and selecting actions like web searching, data analysis, or image creation.

While the integration of LLM’s APIs and the development of plugins and GPTs offer significant advantages, they also present several challenges, such as:
(1) Managing the computational and financial costs associated with training, fine-tuning, and deploying LLM models, which can be significantly higher than those for traditional software systems.
(2) Ensuring that LLM usage adheres to ethical guidelines, particularly in areas like content generation and decision-making.
(3) Verifying the accuracy and reliability of the output generated by LLMs to ensure that plugins and GPTs meet development goals and user expectations.

\section{Methodology}
\label{sec:methodology}

\begin{figure}[htbp]
    \centering
    \includegraphics[width=0.85\textwidth]{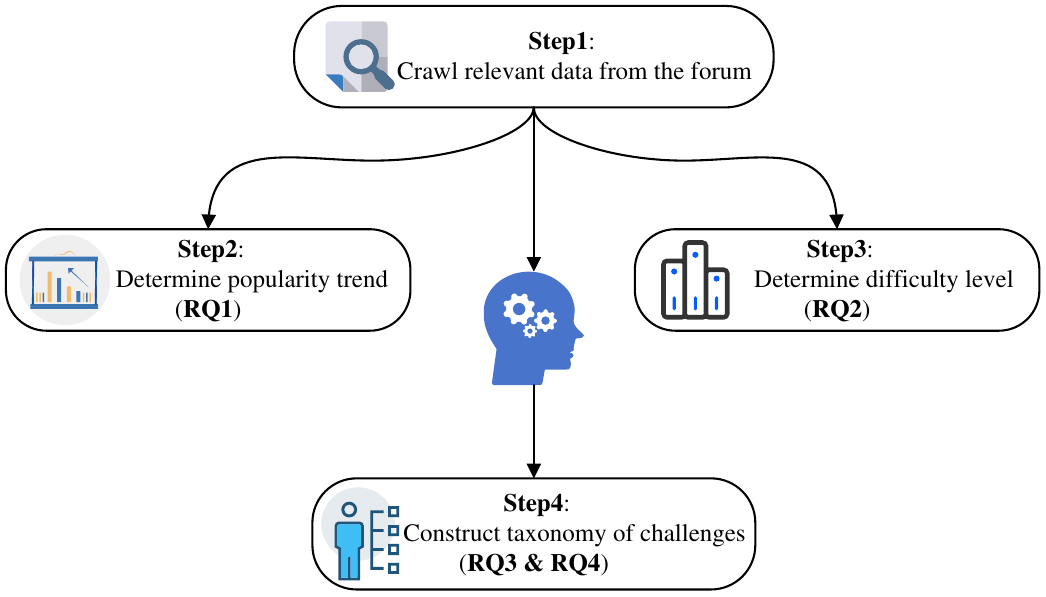} 
    \caption{Methodology overview of our empirical study.} 
    \label{fig:methodology} 
\end{figure}

To understand the challenges faced by LLM developers, we collect and analyze relevant question posts from the OpenAI developer forum, which is a forum provided by OpenAI for developers working with their LLM products.
We mined this forum by following previous studies~\cite{ahmed2018concurrency,alshangiti2019developing,bagherzadeh2019going,lou2020understanding,rosen2016mobile,yang2016security,zhang2019empirical}. 
Fig.~\ref{fig:methodology} provides a methodology overview of our empirical study.

\textbf{Step 1: Crawl relevant data from the forum.}
To analyze the popularity trends and difficulty levels of the questions, we crawl posts from the OpenAI developer forum. 
To analyze challenges faced by LLM developers, we select related subforums of this forum (each subforum's posts are centered around a specific topic, which makes it easier to find relevant posts and ensures that questions reach the appropriate experts).
Our study focuses on four subforums (i.e., API, Prompting, GPT builders, and ChatGPT) in this forum.
Specifically, posts in the ``API" subforum concern questions, feedback, and best practices around building with OpenAI's product API.
posts in the ``Prompting" subforum learn more about prompting by sharing best practices, and favorite prompts.
GPTs are a new way for anyone to create a tailored version of ChatGPT to be more helpful in their daily life, therefore, the ``GPT builders" subforum is designed for developers who are building GPTs.
Finally, the ``ChatGPT" subforum concerns questions or discussions about ChatGPT.
Since posts in the remaining subforum (i.e., Community, Announcements, Documentation, and Forum feedback) rarely discuss the challenges faced by LLM developers, we ignore these posts in our empirical study. 
Specifically, posts in the ``Announcements" subforum typically contain official product updates, posts in the ``Documentation" subforum offer tutorials and instructional content, posts in the ``Community" subforum involve project and technology sharing, and posts in the ``Forum feedback" subforum focus on improving the developer forum.

\begin{figure}[htbp] 
    \centering 
    \includegraphics[width=0.9\textwidth]{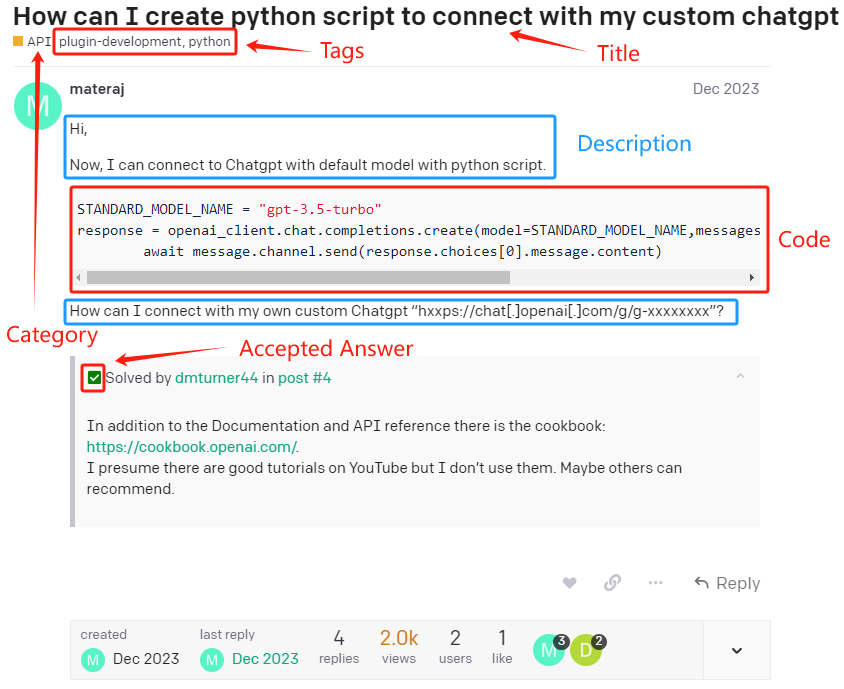} 
    \caption{A question post from the OpenAI developer forum.} 
    \label{Fig:posts_data} 
\end{figure}

Fig.~\ref{Fig:posts_data} shows a screenshot of a post\footnote{\url{https://community.openai.com/t/547685}} from the OpenAI developer forum. Typically, a post consists of several key elements: title, category, tags, question description, code snippet, and the accepted answer (These elements have been marked in this figure). These posts also include additional relevant information, such as creation time, reply time, reply count, view count, and the number of participating users. In this step, we gather the following metadata: the post's title, creation date, the reply time, and the number of replies. We also collect user metadata including usernames and registration times. In summary, by June 2024, we had crawled data from {\postnum} posts and {\usernum} users.

\textbf{Step 2: Determining popularity trend.}
To analyze the popularity trend of LLM development among developers, we first perform the time series analysis. Specifically, we count the number of new posts and new users over different periods (i.e., every three months) from the forum's creation in February 2021 until our data collection deadline in June 2024. Detailed result analysis can be found in Section~\ref{sec:popularity}.

\textbf{Step 3: Determining difficulty level.}
Questions related to LLM development often present significant challenges when attempting to solve them. To better assess the difficulty of these questions, we analyze them from three key perspectives by following the previous studies on Stack Overflow mining~\cite{abdellatif2020challenges,rosen2016mobile,alshangiti2019developing}: reply numbers, accepted solution ratio, and the time required to receive the first reply. Specifically, the number of replies helps gauge community engagement, the accepted solution ratio indicates how frequently a question reaches a satisfactory resolution, and the time required to receive the first reply provides insight into how quickly developers engage with the question. By combining these perspectives, we aim to capture a more holistic view of the difficulty level of LLM-related questions. Detailed result analysis can be found in Section~\ref{sec:difficult}.

\textbf{Step 4: Constructing the taxonomy of challenges.}
From the {\postnum} questions, we randomly selected a statistically significant sample to avoid the high cost of manually analyzing all the crawled posts. This sample size ensures a 99\% confidence level with a $\pm$2.5\% confidence interval~\cite{aghajani2019software}. As a result, {\samplednum} questions constitute the taxonomy dataset of our study. We show the taxonomy construction methodology as follows.

First, we randomly select 70\% of the {\samplednum} posts to construct the initial taxonomy of challenges by following the previous study~\cite{wen2021empirical}. This process involves two annotators working in a collaborative way. We use the open coding procedure~\cite{seaman1999qualitative} to analyze the sampled posts, summarize the preliminary taxonomy, and construct a multi-level hierarchical taxonomy of challenges faced by LLM developers. Each annotator has two years of experience in OpenAI development and three years of experience in developing large language models. In this procedure, they need to carefully read the questions of the sampled posts to understand the specific challenges faced by LLM developers and the corresponding replies. Specifically, the open coding procedure is conducted as follows: the two annotators read all the sampled posts, with each post being read at least twice. During the analysis, they examine all the data contained in each post, including titles, problem descriptions, code snippets, and replies, summarizing the specific challenges discussed.
Then they summarize the specific challenges discussed in the posts. Specifically, some posts describe issues encountered during the use of ChatGPT, for example, ``\emph{Can’t log in to Chat}"\footnote{\url{https://community.openai.com/t/23899}}, which are not closely related to developers.
In addition, there are also some questions related to account problems, tips for users, etc. These posts are labeled as ``Unrelated". Since the OpenAI developer forum is a platform for all OpenAI users to communicate, some posts are official announcements, document introductions, project discussions, and project sharing. For example, ``\emph{I have an idea about creating a mix of emoji}"\footnote{\url{https://community.openai.com/t/22180}}, these posts are temporarily labeled as ``Unrelated". If a post is labeled as ``Unrelated", it means that these posts are not considered when constructing the taxonomy because they are not closely related to LLM developers. For the remaining posts, the annotators use simple phrases to summarize the developer's challenges during the first reading. Specifically, some problem descriptions are about developers consulting on parameter settings when calling OpenAI's APIs, for example, ``\emph{GPT-4-Vision-Preview fidelity/detail-parameter}"\footnote{\url{https://community.openai.com/t/477563}}, the annotators find that this problem is asked because developers do not know how to set API parameters. Therefore, for these issues, the annotators can temporarily describe them as ``API Usage Issue, Parameter Setting".

The two annotators continue to process all sampled posts according to this step and construct the taxonomy of challenges faced by LLM developers. The grouping process is iterative, with continuous modifications to the taxonomy based on the posts and optimization of the post descriptions to improve taxonomy quality. If a post involves multiple categories, it is assigned to all relevant categories. When the two annotators have different opinions on category determination, a third arbitrator is asked to coordinate conflicts. The third arbitrator has five years of software project development experience and three years of experience in developing large language models. After rigorous processing, all participants reach a consensus, and the challenges of all posts are classified, resulting in a preliminary taxonomy of challenges faced by LLM developers.

Finally, we refine the taxonomy of challenges and conduct a reliability analysis. Specifically, the two annotators independently label the remaining 30\% of the posts based on the generated preliminary taxonomy. They label each post as ``Unrelated" or assign it to the corresponding category. If some posts can not be categorized, they are temporarily labeled as ``Pending".
We use Cohen’s Kappa (\emph{k})~\cite{cohen1960coefficient} to measure the consistency between the two annotators during the independent labeling process. The computed $k$ value is 0.812, indicating that the two annotators almost completely agree~\cite{landis1977measurement}. This demonstrates the reliability of our open coding procedure. 
In addition, for posts labeled as ``Pending", we introduce the third arbitrator who discusses and analyzes the posts with the two annotators to determine which category these posts ultimately belong to. 
If they can not be categorized into the preliminary taxonomy, new categories are added. In this phase, we incorporate five additional leaf categories into the preliminary taxonomy framework. Finally, our constructed taxonomy covers all posts related to OpenAI development, and all participants agree on the classification. Detailed result analysis can be found in Section~\ref{sec:taxonomy}. Finally, we also apply this methodology to analyze the challenges of LLM application developers by mining GitHub issues related to more LLMs and the detailed results can be found in Section~\ref{sec:github}.

\section{RQ1: Popularity Trend Analysis}
\label{sec:popularity}

\begin{figure}[htbp] 
    \centering
    \includegraphics[width=0.88\textwidth]{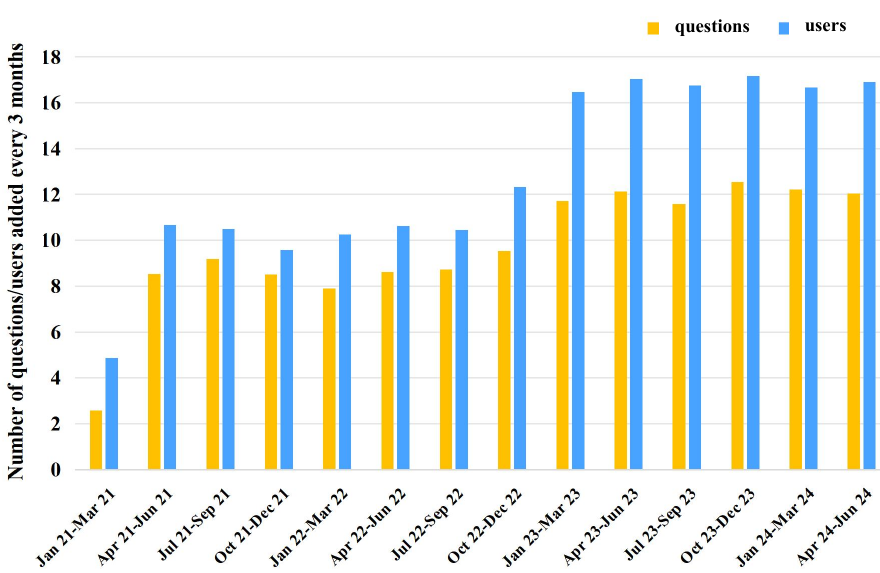} 
    \caption{The number of new posts and users every three months since creating the OpenAI developer forum. Notice, in the $y$-axis, we apply a logarithmic transformation to the actual numbers for a clearer representation of the results}
    \label{Fig:RQ1}
\end{figure}

Fig.~\ref{Fig:RQ1} illustrates the rising trend in LLM development popularity, as evidenced by the increasing number of posts on the OpenAI forum and the surge in registered users. The figure begins at the forum's creation, tracking the growth in new posts and users every three months. Given that the user number significantly exceeds the post number, we apply a logarithmic transformation to the actual numbers for a clearer representation of the results. This visualization reveals a significant increase in developer interest in OpenAI since 2021, underscoring the timeliness and urgency of our empirical study on challenges for LLM application developers.

Before November 2022, the growth trend was fairly consistent. However, the introduction of GPT-3.5 and its fine-tuned product, ChatGPT, to the public by OpenAI in late November 2022, marked a substantial shift. The number of posts increased by 292\%, and the number of users rose by 1,302\% compared to the preceding period. The release of GPT-4 to paying customers in March 2023 further accelerated this trend, with a 323\% increase in posts and a 628\% increase in users. This strong growth trend has persisted until now. Regarding the results of RQ1, we believe that this trend is not just good news for the LLM vendors (such as OpenAI), but has also driven the increasing adoption of LLMs within the software engineering community~\cite{fan2023large,wang2024software} and leads to the emergence of the LLM4SE (LLMs for Software Engineering) research field. The increasing use of LLMs suggests that their application is transforming software development, testing, and maintenance processes, especially in tasks such as code generation~\cite{yan2023closer,gu2024effectiveness}, code completion~\cite{eghbali2024hallucinator}, test generation~\cite{schafer2023empirical,yuan2023no}, program repair~\cite{jin2023inferfix}, vulnerability detection~\cite{lu2024grace,steenhoek2024dataflow}, and source code summarization~\cite{sun2024source} (The latest research progress for these tasks can be referenced in Section~\ref{sec:relatedLLM4SE}). This trend highlights the significant role LLMs can play in enhancing productivity and efficiency across various stages of the software engineering lifecycle.

\finding{1}{
\textbf{Finding 1.} The introduction of GPT-3.5 and ChatGPT in November 2022, followed by GPT-4 in March 2023, significantly accelerated the growth of the LLM developer community, markedly increasing the number of posts and users on the OpenAI developer forum.}

\section{RQ2: Difficulty Level Analysis}
\label{sec:difficult}

As described in Section~\ref{sec:methodology}, we evaluate the difficulty level of LLM-related questions from three different perspectives.

\begin{figure}[htbp] 
    \centering 
    \includegraphics[width=0.8\textwidth]{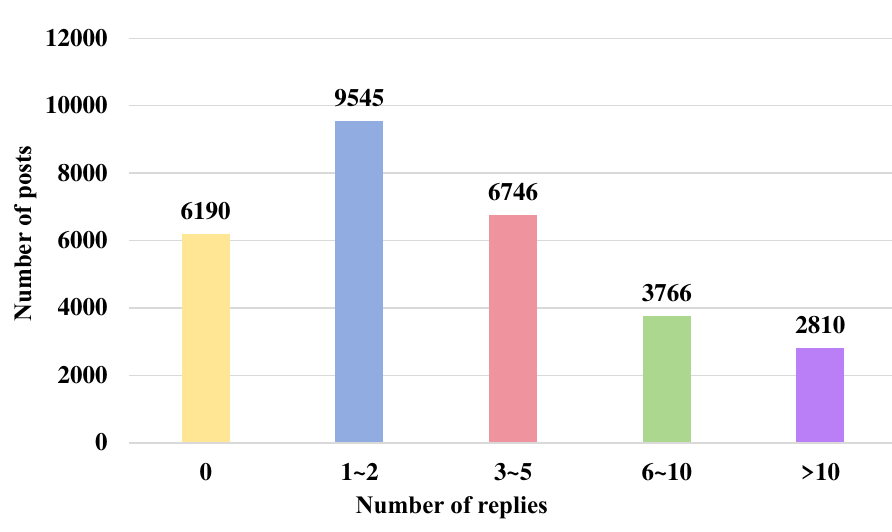} 
    \caption{The number of posts with different numbers of replies.} 
    \label{Fig:RQ2} 
\end{figure}

First, we use Fig.~\ref{Fig:RQ2} to show the number of posts with different numbers of replies. The results show that 54\% of questions receive fewer than three replies, while only 10\% receive more than ten replies. This suggests that questions related to LLM development are generally challenging to resolve. However, we acknowledge that using the number of replies as a metric for assessing question difficulty has its limitations. For example, a greater number of responses to a question might suggest that reaching an agreement requires discussions from multiple developers or several iterations. On the other hand, questions with fewer replies could be simpler to solve, leading to less discussion.

\begin{figure}[htbp] 
    \centering 
    \includegraphics[width=0.45\textwidth]{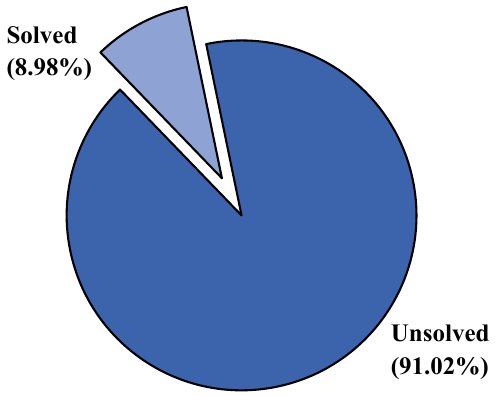} 
    \caption{Proportion of posts with accepted solutions. } 
    \label{Fig:RQ2pro} 
\end{figure}

Then, we examine the proportion of posts with an accepted solution. Fig.~\ref{Fig:RQ2pro} illustrates the percentage of posts that have an accepted solution. From these statistical results, we find that only 8.98\% of the posts have accepted solutions. The relatively low acceptance rate suggests that these questions are generally more difficult to resolve. Moreover, a low acceptance rate may indicate that the quality assurance mechanism of the community's posts is relatively immature compared to more established developer Q\&A platforms like Stack Overflow, where accepted solutions are marked more frequently. 

\begin{figure}[htbp] 
    \centering 
    \includegraphics[width=0.75\textwidth]{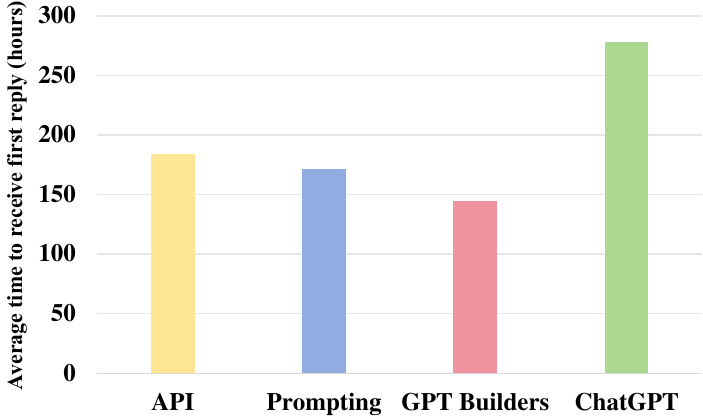} 
    \caption{The average time required to receive the first reply for each subforum. } 
    \label{Fig:RQ2interval} 
\end{figure}

Finally, we analyze the time required to receive the first reply. This metric offers additional insight into the complexity of the questions, as a longer response time may indicate that the question is more difficult to address. Fig.~\ref{Fig:RQ2interval} presents the average response time of questions with the replies for each subforum. As shown in this figure, most posts have a relatively long response time (approximately between 147 hours and 278 hours for different subforums), suggesting that many of these questions are challenging or that they require more time for developers to give a response.

Based on the above results from three different perspectives, we summarize potential reasons as follows. First, the new and complex technological domains of LLM require deep expertise and skills from developers. This scarcity of qualified professionals results in limited responses. 
Second, the rapid evolution of LLM technology often leaves issues unresolved, leading to fewer available answers. 
Finally, the lack of comprehensive documentation and tailored support resources makes it difficult to address diverse developer needs, prolonging the resolution process for many questions.

\finding{2}{
\textbf{Finding 2.} Problems related to LLM development often face challenges in receiving immediate, sufficient, and even accepted replies. This highlights potential directions for the community and LLM vendors to offer more effective support.}

\section{RQ3: Challenge Taxonomy Construction}
\label{sec:taxonomy}

\begin{figure}[htbp] 
    \centering
    \includegraphics[width=0.96\textwidth]{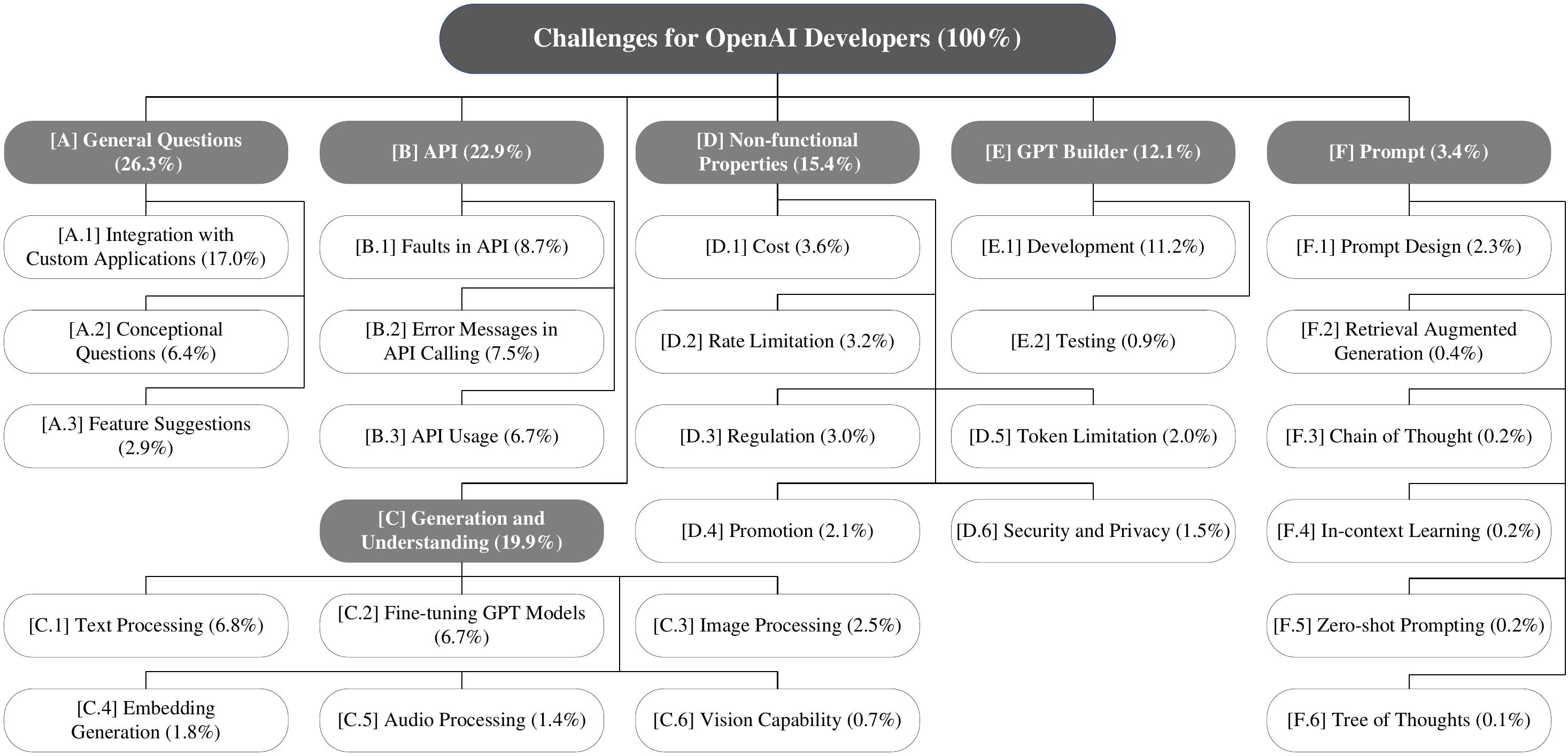} 
    \caption{Our Constructed Challenge Taxonomy for LLM Developers.} 
    \label{Fig:category_important}
\end{figure}

Fig.~\ref{Fig:category_important} illustrates the hierarchical taxonomy of challenges faced by LLM developers. In this figure, the nodes are shaded in descending levels of grey based on their depth in the hierarchy (notice leaf nodes are white). Each leaf node represents a specific category, while its parent node is an inner node composed of multiple subcategories. For example, \emph{API (B)} is an inner category that can be further classified into three leaf categories: \emph{Faults in API (B.1)}, \emph{Error Messages in API Calling (B.2)}, and \emph{API Usage (B.3)}.

The proportion of posts for each category is shown in parentheses. In summary, our taxonomy includes six inner categories and 26 leaf categories. Based on this taxonomy, we find that LLM developers encounter a wide variety of problems across different aspects of development, highlighting the diversity of challenges in LLM development. 
Next, we describe and provide examples for each category and summarize our findings and implications for developers and LLM vendors.

\finding{3}{
\textbf{Finding 3.} The challenges faced by LLM developers are multifaceted and diverse, encompassing 6 categories and 26 distinct subcategories.}

\subsection{General Questions}
 
General Questions encompass challenges that lack specific implementation details and are often posed by developers seeking foundational knowledge in LLM development. These challenges reflect the diverse issues developers face when starting with or working on LLM-based applications. Common difficulties include understanding how to integrate LLMs with custom applications, addressing conceptual gaps (e.g., token usage and API functionalities), and suggesting enhancements to existing features. 
Our findings indicate that 26.3\% of the challenges fall into this category. The detailed subcategories are discussed as follows.

\begin{figure}[htbp] 
    \centering 
    \includegraphics[width=0.85\textwidth]{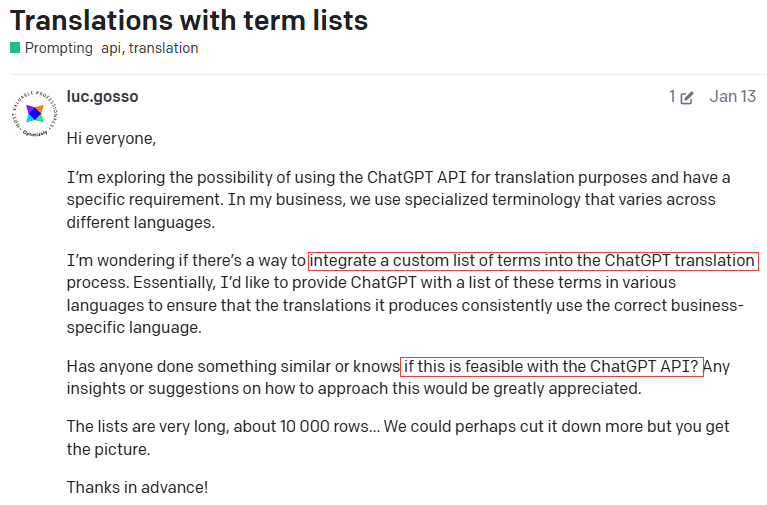} 
    \caption{A sample post in the integration with custom applications (A.1) related to specialized terminology.} 
    \label{Fig:post1} 
\end{figure}

\begin{figure}[htbp] 
    \centering 
    \includegraphics[width=0.85\textwidth]{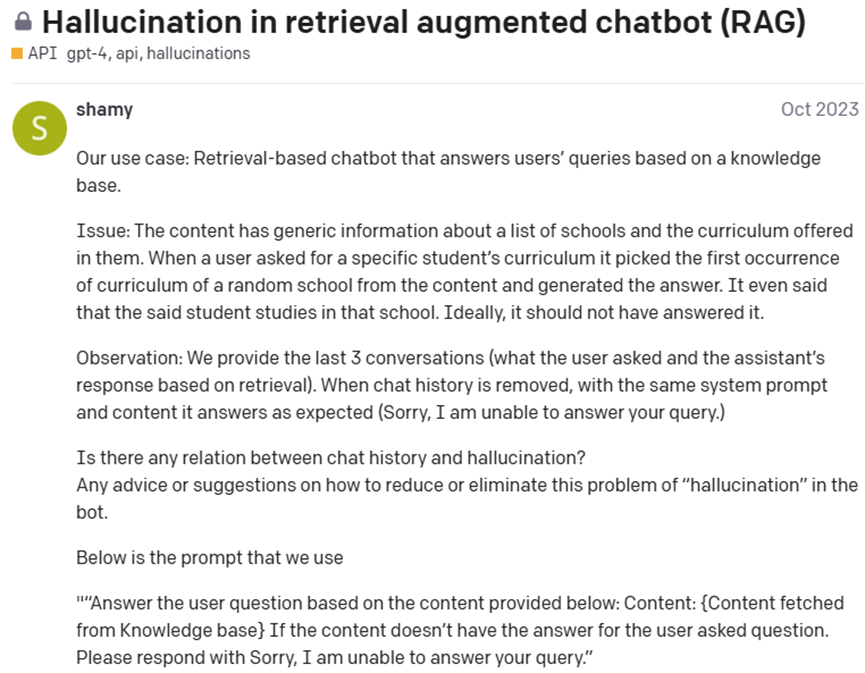} 
    \caption{A sample post in the integration with custom applications (A.1) related to hallucinations.} 
    \label{Fig:post11} 
\end{figure}

\textbf{Integration with Custom Applications (A.1).}
With the launch of ChatGPT and the availability of LLM's APIs, more developers are leveraging these GPTs/APIs to build custom applications, such as translation systems and text classification systems. Similar to the challenges in ML application integration~\cite{alshangiti2019developing}, we find that developers integrating LLMs with their applications also encounter difficulties, such as API configuration, environment setup, and the outputs of LLMs.
For example, developers seek to understand how to integrate custom GPTs with web page interfaces\footnote{\url{https://community.openai.com/t/589076}} or inquire about the feasibility of implementing specific functionalities through API calls, such as
considering specialized terminology that varies across different languages\footnote{\url{https://community.openai.com/t/584762}}, as shown in Fig.~\ref{Fig:post1}. 
Additionally, some developers raise concerns about the reproducibility of results in commercial text classification systems using GPT-3.5-turbo, especially after model updates\footnote{\url{https://community.openai.com/t/281658}}. Developers also encounter problems related to hallucinations\gcy{\footnote{\url{https://community.openai.com/t/408275}}}, as shown in Fig.~\ref{Fig:post11}.
Our findings indicate that challenges in this subcategory account for a significant 17.0\%, highlighting the diverse and complex technical issues developers face when building their custom applications.

\textbf{Conceptional Questions (A.2).}
This subcategory encompasses questions about basic concepts or background knowledge of LLM development, such as how to calculate the tokens needed for each question\footnote{\url{https://community.openai.com/t/81018}}, whether the GPT-4 API has access to the Internet\footnote{\url{https://community.openai.com/t/468615}}, or whether plugins can be used with the API\footnote{\url{https://community.openai.com/t/186333}}. 
Additionally, developers inquire about the internal implementation details, such as mechanisms of the content filter\footnote{\url{https://community.openai.com/t/1095}}. 
In our taxonomy, the posts in this subcategory account for 6.4\%.

\textbf{Feature Suggestions (A.3).}
Developers often suggest adding new features related to API operations or express a desire to develop plugins for specific functionalities. 
For example, developers want to provide a way to view historical usage and spending by individual API key for better cost management of deployed chatbot apps using OpenAI's GPT turbo API\footnote{\url{https://community.openai.com/t/305606}}. 
Additionally, they recommend adding search and categorization features to the plugin page, which can quickly find the desired plugins\footnote{\url{https://community.openai.com/t/215754}}. 
Challenges related to these suggestions account for 2.9\% in our taxonomy.

\finding{4}{
\textbf{Finding 4.} 
The majority (26.3\%) of the total challenges fall under the category of \emph{General Questions}. Within this category, the subcategory \emph{Integration with Custom Applications} represents the largest proportion, accounting for 64.6\% of \emph{General Questions}.
}

\textbf{Discussion and Implication:} 
General questions often reflect fundamental challenges developers face when working with LLM technologies. These challenges span various aspects (such as application integration, conceptual questions, and feature suggestions). 
Similar to the empirical studies~\cite{zhang2019empirical} on common challenges in developing deep learning applications,
when leveraging APIs provided by OpenAI to build their applications (such as translation systems, text analysis systems, speech recognition systems, and others), developers often encounter complex technical questions. These include API integration methods, performance issues, output reproducibility issues, interpretability of output content, and so on. These challenges highlight the diverse and intricate nature of API integration and the need for clearer guidelines and examples from the LLM vendors to assist developers in these areas. 
As integration with custom applications is a major challenge, The LLM vendors could develop dedicated resources and support mechanisms to streamline this process. This may include detailed tutorials and integration best practices.
Questions in the subcategory of conceptual questions involve basic concepts and background knowledge necessary for understanding OpenAI’s technologies. Developers often inquire about token calculations, API capabilities and internal mechanisms, model versions, and other related issues. These challenges underscore the need for improved foundational documentation. To mitigate these challenges, as suggested by Chen et al.~\cite{chen2020comprehensive}, improving the usability and completeness of the documentation will help developers grasp fundamental concepts as their required skills more effectively, reducing the frequency of basic questions and accelerating their development process.
For the last subcategory, developers frequently suggest new features or improvements related to API operations, such as better cost management tools, enhanced plugin and GPTs recommendation features, and support for a wide range of data formats. This subcategory reflects developers' proactive interest in enhancing OpenAI’s tools. Therefore, the LLM vendors (e.g., the OpenAI organization) should establish a structured feedback mechanism to capture and address feature suggestions from developers. This will help prioritize feature development based on actual user needs and enhance overall user satisfaction.

\subsection{API}
This category contains challenges related to the LLM APIs, including issues with fault handling, error messages, and parameter configurations, which collectively hinder development efficiency and create barriers to effective usage. Common challenges include managing unsatisfactory outputs, troubleshooting error messages, and understanding or optimizing API parameters for specific tasks. These challenges highlight the complexity of integrating LLM APIs into practical applications and underscore the need for clearer documentation and support mechanisms. 
In our taxonomy, 22.9\% of the total challenges fall under this category.
The detailed subcategories are discussed as follows.

\textbf{Faults in API (B.1).}
When calling LLM APIs, developers frequently encounter a variety of issues, such as low-quality generated content, limitations in model comprehension, and text coherence problems. These issues often result in LLM outcomes that do not meet developers' expectations.  The majority of these issues are related to unsatisfactory output, such as the presence of extraneous information (like spaces and newlines) in the API's responses\footnote{\url{https://community.openai.com/t/578701}}, as well as phrase repetition in answers\footnote{\url{https://community.openai.com/t/54737}}. 
Beyond these common faults, developers face other types of faults, such as APIs failing to respond due to requests being too frequent\footnote{\url{https://community.openai.com/t/483090}}. Additionally, when dealing with specific technical problems, such as converting natural language into SQL queries, they might produce incorrect answers\footnote{\url{https://community.openai.com/t/426293}}.
These examples illustrate the diverse technical challenges that can arise. This subcategory represents 8.7\% of the total challenges.

\begin{figure}[htbp] 
    \centering 
    \includegraphics[width=0.88\textwidth]{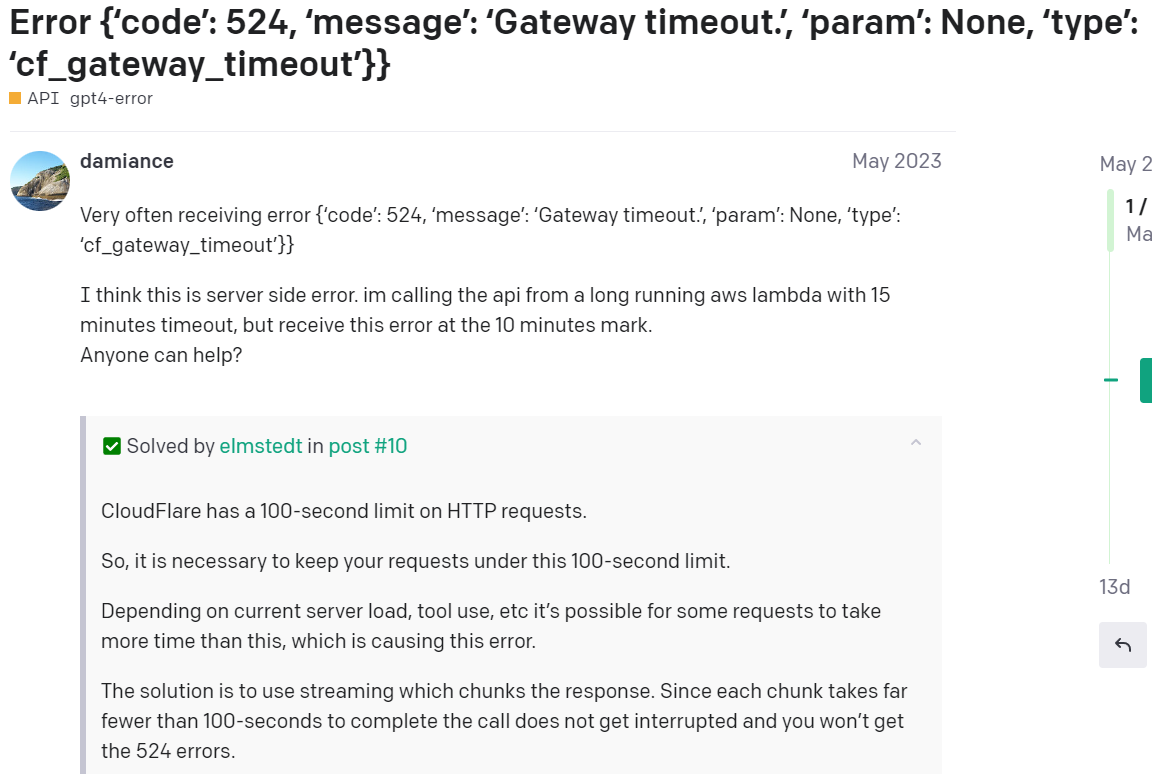} 
    \caption{A sample post in the error messages in API calling (B.2).} 
    \label{Fig:post2} 
\end{figure}

\textbf{Error Messages in API Calling (B.2).}
This category primarily includes various error messages that developers encounter when calling LLM APIs, which is also faced by DL application developers~\cite{morovati2024common}. For example, developers might face request errors\footnote{\url{https://community.openai.com/t/154711}} and data value capacity limit errors when calling an API for image editing\footnote{\url{https://community.openai.com/t/464513}}. As shown in Fig.~\ref{Fig:post2}, developers may also face issues related to the OpenAI server, such as gateway timeout errors\footnote{\url{https://community.openai.com/t/219369}}. This subcategory represents 7.5\% of the total challenges.

\begin{figure}[htbp] 
    \centering 
    \includegraphics[width=0.9\textwidth]{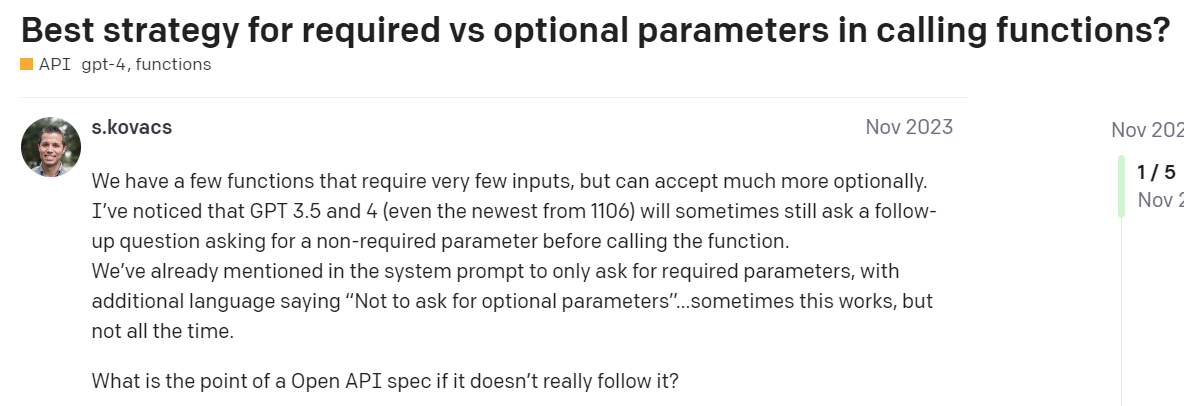} 
    \caption{A sample post in API usage (B.3).} 
    \label{Fig:post3} 
\end{figure}

\textbf{API Usage (B.3).} 
Similar to challenges in developing deep learning applications~\cite{zhang2019empirical,morovati2024common}, the parameter configuration is a principal concern among LLM developers, necessitating the parameter value adjustment to meet specific requirements and generate expected outputs from the API. 
A significant portion of these problems center on determining the appropriate values for API parameters\footnote{\url{https://community.openai.com/t/480243}} (as shown in Fig.~\ref{Fig:post3}) and understanding the internal working and implementation mechanism (such as the \emph{response\_model} parameter\footnote{\url{https://community.openai.com/t/511738}}). Additionally, some challenges are closely associated with GPT models. For example, setting identical values for the seed parameter does not result in consistent outputs for specific GPT models\footnote{\url{https://community.openai.com/t/487245}}. This subcategory represents 6.7\% of the total challenges.

\finding{5}{
\textbf{Finding 5.} 
Challenges related to the \emph{API} rank second, accounting for 22.9\% of the total challenges. Within this category, \emph{Faults in API} and \emph{Error Messages in API Calling} represent 38.0\% and 32.8\% of the challenges, respectively.
}

\textbf{Discussion and Implication:} 
The use of APIs is a challenge faced by developers in many fields, such as desktop applications~\cite{scoccia2021challenges}, web applications~\cite{samudio2022barriers}, and deep learning applications~\cite{cao2022understanding,morovati2024common}.
Our results show that developers also frequently encounter various issues when using OpenAI APIs, leading to a decline in development efficiency. Some of these problems are related to the specific GPT models being used. For example, certain APIs only support the gpt-4 turbo model for specific fields like response\_format, indicating potential compatibility issues. To mitigate confusion and reduce the frequency of these issues,
the OpenAI organization should ensure that their documentation clearly outlines the compatibility of different models and fields, and provide detailed examples and best practices for developers to follow. Moreover, researchers can develop tools for detecting and repairing these compatibility issues by following the suggestions of Wang et al.~\cite{wang2023compatibility} after analyzing compatibility issues in DL systems.
Developers also encounter various error messages when calling OpenAI API.  To address these issues, the OpenAI organization should consider providing more detailed error messages that help developers understand the cause of the errors and how to fix them. Additionally, offering guidance on common parameter configurations and their implications can be beneficial. For example, addressing phrase repetition issues can be managed by adjusting the frequency\_penalty parameter, where setting positive values can penalize new tokens based on their existing frequency in the text so far, thereby decreasing the model's likelihood of repeating the same line verbatim.
Parameter configuration is a principal concern among developers. To help developers, the OpenAI organization should provide comprehensive documentation that includes detailed explanations of parameter values and their effects. Offering sample configurations and use-case scenarios can guide developers in making informed decisions about parameter settings. For example, explaining how different temperature parameter values affect the generated content's creativity and randomness can help developers tailor the API output to their needs.
In summary, LLM vendors should consider five important factors when designing API documentation (i.e., documentation of intent, code examples, matching APIs with scenarios, the penetrability of the API, and format and presentation), as suggested by Robillard and Deline~\cite{robillard2011field} and improve API usability based on human-centered design methods (such as understanding how developers think about API functionality to design more intuitive interface, ensuring consistent parameter order and naming conventions, and avoiding long, confusing parameter lists), as suggested by Myers and Stylos~\cite{myers2016improving}. Moreover, providing more informative error messages can help developers quickly identify and resolve issues. Finally, they should monitor community forums and address recurring issues by updating the API and documentation accordingly as suggested by Chen et al. when analyzing challenges in deploying deep learning-based software~\cite{chen2020comprehensive}.

\subsection{Generation and Understanding}
Due to powerful generation and understanding capabilities, developers can use LLMs to perform a variety of tasks (such as text processing via the LLM GPT-4o, image processing via the LLM DALL-E, and audio processing via the LLM Whisper) or fine-tune the OpenAI LLMs for downstream tasks. 
However, developers frequently face challenges in effectively leveraging these capabilities. Common issues include troubleshooting errors in text, image, audio, and embedding processing tasks; managing fine-tuning processes; and addressing API-related limitations. These challenges often involve optimizing input prompts, resolving parameter configuration problems, and addressing unexpected API errors or quality inconsistencies in outputs. 
Notice the challenges in these tasks often overlap with other challenge categories, so a question (such as parameter setting issues related to image generation\footnote{\url{https://community.openai.com/t/359438}}) may be classified into multiple categories. Challenges related to this category account for 19.9\% of the total challenges. 
The detailed subcategories are discussed as
follows.

\textbf{Text Processing (C.1).}  
As LLM's core capability, text generation helps developers to create coherent and semantically sound textual content based on the input text (i.e., the prompt). This feature allows developers to automatically generate various types of text (such as articles, stories, and dialogues). By modifying the input prompts, developers can control the style, content, and length of the generated text to meet diverse application requirements across different scenarios. This category contains challenges associated with text generation, such as encountering errors when using the text generation API\footnote{\url{https://community.openai.com/t/486260}} or getting repeated responses from the LLM davinci\footnote{\url{https://community.openai.com/t/254241}}. Moreover, developers want to provide feedback on the given output to improve the upcoming outputs when using the gpt-3.5-turbo model\footnote{\url{https://community.openai.com/t/452220}}. This subcategory constitutes 6.8\% of all challenges.

\begin{figure}[htbp] 
    \centering 
    \includegraphics[width=0.85\textwidth]{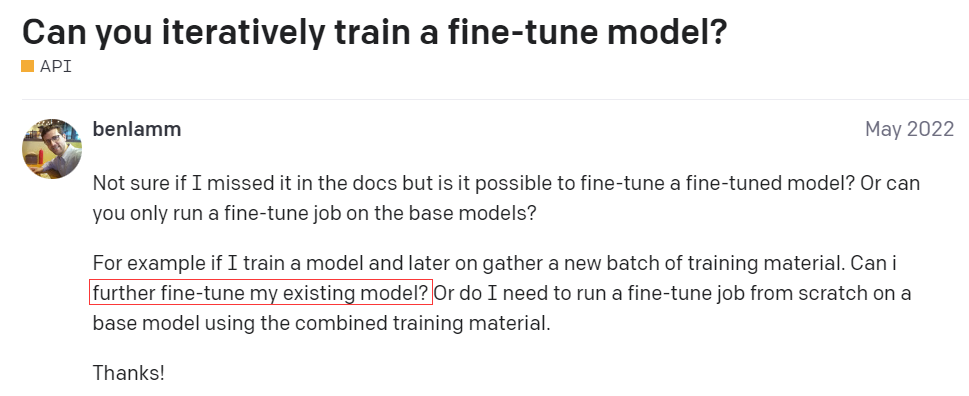} 
    \caption{A sample post in fine-tuning GPT models (C.2).} 
    \label{Fig:post8} 
\end{figure}

\textbf{Fine-tuning GPT Models (C.2).}
LLM developers can fine-tune the models provided by the OpenAI organization by using the training datasets of the downstream tasks. Due to the widespread use of fine-tuning, developers frequently face questions related to this functionality. For example, they often inquire about how to set the parameter \emph{n\_epochs}\footnote{\url{https://community.openai.com/t/339246}} or request the fine-tuned model to output results in JSON format\footnote{\url{https://community.openai.com/t/525169}}. Additionally, as shown in Fig.~\ref{Fig:post8}, developers are interested in whether they can train a fine-tuned model in an iterative way\footnote{\url{https://community.openai.com/t/18248}}. This subcategory of questions accounts for 6.7\% of the total challenges.

\textbf{Image Processing (C.3).}
Developers can influence image generation by inputting prompts into the model, allowing for the creation of images with customized features and styles. This versatile feature is useful across diverse applications, including image editing and artistic creation, offering developers opportunities for creative expression and design. However, developers often encounter challenges such as request errors when seeking image generation solutions\footnote{\url{https://community.openai.com/t/586006}} or perceiving that the quality of the generated images is lower than expected\footnote{\url{https://community.openai.com/t/435187}}. Finally, they find that the image quality generated through API calls is inferior to that of the ChatGPT web version using the same prompts\footnote{\url{https://community.openai.com/t/435187}}. This subcategory constitutes 2.5\% of all challenges.

\textbf{Embedding Generation (C.4).}
By using the Embedding APIs provided by LLM vendors, developers can leverage powerful text representation capabilities to convert text into high-dimensional vector representations. These vectors can capture the semantics and contextual information of the text, which can be used for tasks such as semantic search and text clustering. For example, developers may encounter \emph{PermissionError} when creating embeddings\footnote{\url{https://community.openai.com/t/587579}}. Additionally, they often inquire whether modifying user queries can improve the accuracy of semantic search\footnote{\url{https://community.openai.com/t/393047}} and how to determine the length of embedded content\footnote{\url{https://community.openai.com/t/111471}}. Challenges related to this subcategory account for 1.8\% of the total challenges.

\begin{figure}[htbp] 
    \centering 
    \includegraphics[width=0.8\textwidth]{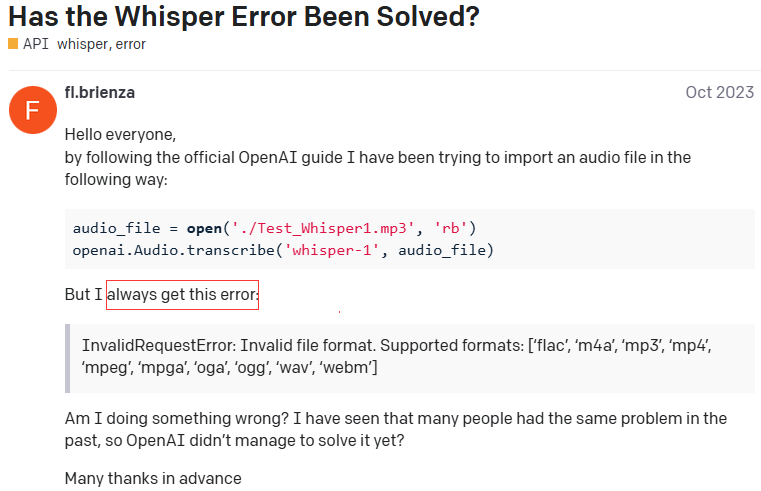} 
    \caption{A sample post in audio processing (C.5).} 
    \label{Fig:post4} 
\end{figure}

\textbf{Audio Processing (C.5).}
Processing audio is one of the key functionalities provided by the OpenAI organization. Developers can utilize advanced speech models (such as the LLM Whisper) to perform transformations, analysis, and generation of audio data, thereby supporting applications such as speech recognition and audio generation. For instance, developers might encounter random text in the output after the API recognizes audio\footnote{\url{https://community.openai.com/t/287544}} or face \emph{InvalidRequestError} when calling the LLM whisper API\footnote{\url{https://community.openai.com/t/433315}}, as shown in Fig.~\ref{Fig:post4}. Additionally, they often ask how to combine audio processing with image processing\footnote{\url{https://community.openai.com/t/22912}}. Challenges in this subcategory constitute 1.4\% of all challenges.

\textbf{Vision Capability (C.6).} 
The Vision API uses pre-trained vision LLMs to quickly and accurately identify objects, scenes, and features in images, providing developers with powerful image processing capabilities. This functionality is widely applicable in areas (such as image recognition, intelligent surveillance, and medical image analysis). However, developers may encounter error messages when using the gpt-4-vision-preview API\footnote{\url{https://community.openai.com/t/546593}} or find that it occasionally misinterprets information in images\footnote{\url{https://community.openai.com/t/504043}}. They also ask usage-related questions, such as how to add a parameter to view image resolution\footnote{\url{https://community.openai.com/t/477563}}. This subcategory constitutes 0.7\% of all challenges.

\finding{7}{
\textbf{Finding 7.} The category of \emph{Generation and Understanding}, the third-largest category, has 34.2\% of challenges concentrated in \emph{Text Processing} and accounts for 19.9\% of all challenges.}

\textbf{Discussion and Implication:} 
The analysis of the various subcategories within the ``Generation and Understanding" category reveals several insights into the challenges developers face and the implications for improving OpenAI’s LLMs and their usage.
(1) API usage issues. A significant portion of the challenges are related to the practical use of APIs. Issues such as repeated responses from the LLM Davinci, errors in embedding API usage, and problems with the Whisper API during audio processing are common. Developers often struggle with these API-related problems, seeking solutions and best practices for effective implementation. Therefore, detailed, updated documentation should be provided, especially for frequently encountered issues. For instance, recommending the use of the latest APIs, such as the Chat Completions API instead of older models like Davinci, can help mitigate repeated response issues.
Providing best practices for embedding API usage, such as adding an offset vector to improve accuracy, can enhance developer experience. Similarly, sharing solutions for Whisper API errors, like addressing audio recording and encoding issues, can be beneficial. Moreover, to implement API recommendations related to LLM development, a knowledge graph of relevant LLM APIs can be constructed based on API documentation and discussions from the OpenAI forum as suggested by Cao et al.~\cite{cao2022understanding}. (2) Conceptual challenges. Developers often face conceptual challenges, particularly with fine-tuning GPT models, parameter settings, iterative training, and output formatting, indicating a need for a deeper understanding of these processes. Therefore, creating comprehensive guides and tutorials on fine-tuning, common parameter settings, and iterative training techniques, can address many of these questions.
(3) Application-specific issues. Challenges also arise from specific applications. Taking image processing as an example, developers encounter issues like translating prompts from other languages (such as French) to English, intellectual property concerns with generated images, and improving image recognition accuracy by adjusting contrast or rotation. Therefore, enhancing multilingual support for image processing prompts can help generate the desired images regardless of the input language. Providing clear guidelines on the intellectual property rights of generated images can address developers' concerns and promote more confident use of the technology. Sharing techniques for improving image recognition accuracy, such as contrast adjustment and rotation correction, can help developers achieve better results in their applications.

\subsection{Non-functional Properties}
The importance of non-functional properties cannot be ignored by LLM developers and is closely intertwined with LLM development. 
These properties include aspects such as API call costs, rate limitations, and data security and privacy, which significantly influence the efficiency and usability of LLM-based applications. Developers frequently face challenges such as managing API call costs by optimizing token usage, understanding and working within rate limitations, ensuring compliance with regulatory requirements, handling token limitations for large-scale datasets, and safeguarding data security and privacy during API usage. 
In our taxonomy, 15.4\% of the challenges belong to this category. The detailed subcategories are discussed as follows.

\textbf{Cost (D.1).}
Challenges in this subcategory are related to API call costs. In the context of OpenAI development, costs are determined by the number of tokens utilized in each API call, particularly for generated content. Developers express concerns regarding the calculation of token usage and the cost\footnote{\url{https://community.openai.com/t/496993}}. They seek strategies to reduce costs by minimizing token consumption\footnote{\url{https://community.openai.com/t/525033}}. Responses to these challenges typically involve providing documentation-related information or offering strategies that consider factors such as the chosen model, input content length, and response content length. This subcategory constitutes 3.6\% of all challenges.

\begin{figure}[htbp] 
    \centering 
    \includegraphics[width=0.8\textwidth]{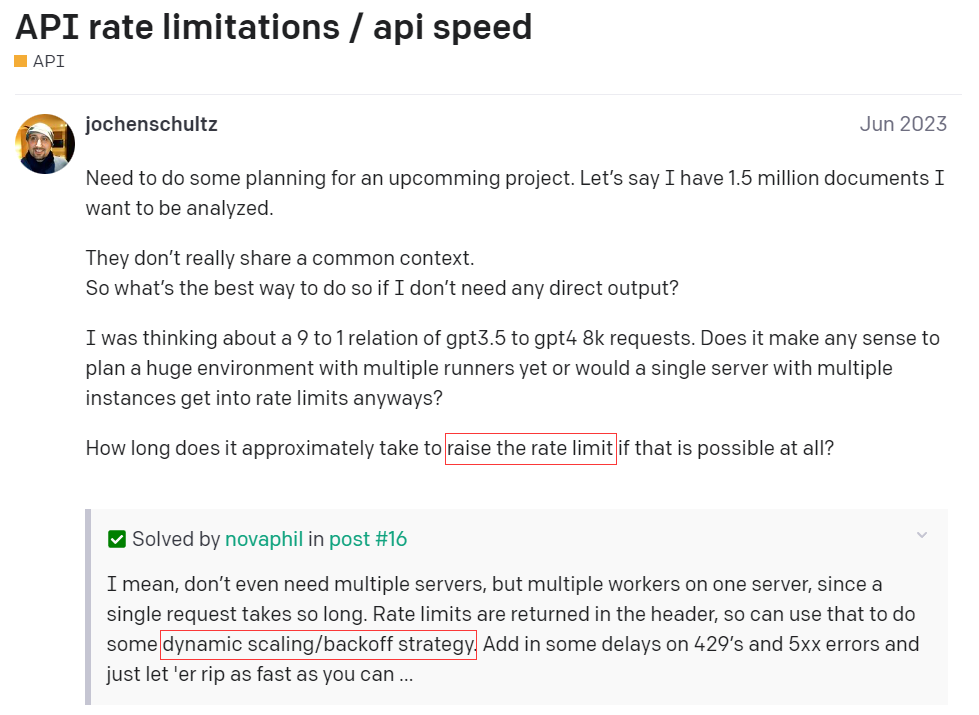} 
    \caption{A sample post in rate limitation (D.2).} 
    \label{Fig:post9} 
\end{figure}

\textbf{Rate Limitation (D.2).}
Rate limitation is typically implemented to safeguard the stability and reliability of OpenAI services, preventing service interruptions or performance degradation due to excessive API usage. Developers need to understand the rate-limiting rules of the API, including the maximum number of calls per second, per minute, or per hour for different versions of the API, and how to manage call frequency in a sensible way to avoid triggering rate limits and causing service interruptions or exceptions. For example, developers encounter \emph{RateLimitError} when calling gpt-3.5-turbo-0301\footnote{\url{https://community.openai.com/t/566696}}. Additionally, developers inquire whether rate limits are shared among different APIs\footnote{\url{https://community.openai.com/t/360331}}. Furthermore, as shown in Fig.~\ref{Fig:post9}, developers ask about methods to alleviate or increase the API rate limits\footnote{\url{https://community.openai.com/t/248374}}. These types of questions account for 3.2\% of the total challenges.

\textbf{Regulation (D.3).}
When utilizing the LLM APIs, developers should strictly adhere to relevant regulations. For instance, the OpenAI organization provides guidelines for the release of specific models or features and determines which users are granted permission to utilize them\footnote{\url{https://openai.com/policies/usage-policies/}}. 
Developers should review and abide by the terms of service and usage policies provided by the OpenAI organization to ensure compliance with commercial and copyright regulations during API calls. For example, developers want to know whether LLM API can be used for commercial purposes\footnote{\url{https://community.openai.com/t/193738}}, or whether the images generated by the LLM DALL-E can be sold\footnote{\url{https://community.openai.com/t/33538}}. Additionally, developers encounter issues where prompts violate content policy rules\footnote{\url{https://community.openai.com/t/468967}}. This subcategory constitutes 3.0\% of all challenges.

\textbf{Promotion (D.4).} 
Developers often leverage forums as platforms to promote their developed plugins and GPTs, aiming to gather user feedback and potentially generate revenue. To this end, they offer comprehensive introductions to their creations, innovative functionalities, and usage guidelines. 
For instance, developers have expressed concerns about the lack of discoverability in the current GPT store, noting that many high-quality GPTs remain largely unknown to most users. To address this issue, they propose implementing a recommendation system to enhance user experience\footnote{\url{https://community.openai.com/t/585652}}.
This category represents 2.1\% of the challenges.

\begin{figure}[htbp] 
    \centering 
    \includegraphics[width=0.85\textwidth]{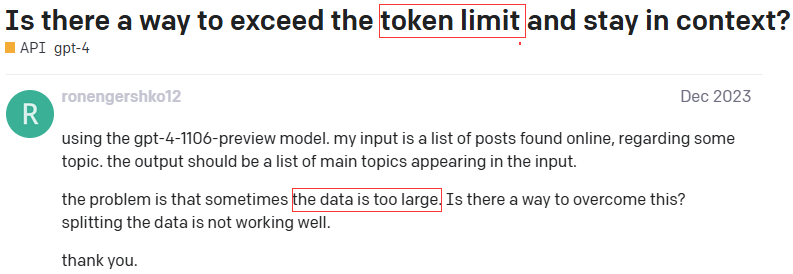} 
    \caption{A sample post in token limitation (D.5).} 
    \label{Fig:post5} 
\end{figure}

\textbf{Token Limitation (D.5).}
In addition to rate limiting, OpenAI imposes restrictions on the number of tokens for input and output during API calls to prevent abuse of API resources. Therefore, developers should understand the allocation, usage rules, and limitations of API tokens, as well as how to manage and optimize their usage. Specifically, developers find that the context length of the API is not fully utilized\footnote{\url{https://community.openai.com/t/291384}}. Additionally, developers seek methods to address token limitation~\footnote{\url{https://community.openai.com/t/4914}}. Furthermore, they inquire about solutions for handling large-scale datasets that exceed the token limitation and ineffective data splitting when using the LLM gpt-4-1106-preview\footnote{\url{https://community.openai.com/t/566336}}, as shown in Fig.~\ref{Fig:post5}. This category constitutes 2.0\% of the challenges.

\textbf{Security and Privacy (D.6).}
Ensuring data security and preventing leakage are primary concerns for LLM developers, especially when using API calls that involve sensitive information. Developers should focus on safeguarding data privacy, adhering to OpenAI's privacy policies, and securely managing API keys to avoid unauthorized access and misuse.
Specifically,
first, ensuring security and preventing data leakage during API calls are the main concerns for LLM developers. 
For example, developers express apprehension about potential data security risks when utilizing the API to process their uploaded files\footnote{\url{https://community.openai.com/t/354485}}. Additionally, they want to know whether using the API within internal networks can lead to data leakage\footnote{\url{https://community.openai.com/t/75725}}. Second, developers are advised to carefully review OpenAI's privacy policy to gain insights into the data collection and processing practices during API calls and to understand the measures in place to protect users' personal privacy. Since API keys serve as critical qualifications for developer authentication and authorization, it is imperative for developers to prioritize the secure storage and management of these keys to prevent leaks or misuse. For example, developers create a desktop application based on ChatGPT and want to know if each user can use their API key within the program for billing purposes\footnote{\url{https://community.openai.com/t/161185}}. Finally, developers have concerns regarding potential privacy issues associated with GPTs\footnote{\url{https://community.openai.com/t/496343}}. This subcategory constitutes 1.5\% of all challenges.

\finding{8}{
\textbf{Finding 8.} 15.4\% of the total challenges are in \emph{Non-functional Properties}, covering seven subcategories. \emph{Cost} and \emph{Rate Limitation} are the top two categories, together accounting for 44.2\% of the challenges in \emph{Non-functional Properties}.}

\textbf{Discussion and Implication:} 
Challenges such as API call costs, rate limitation, and token limitation are tightly linked to the development and usage of the LLM's services. Developers often express concerns about the costs associated with API calls, which are influenced by the choice of model and the number of tokens used in each request. Similarly, rate limitations are put in place to ensure service stability, but developers need to understand these limits and manage their API call frequencies accordingly. Token limitations pose additional challenges, especially when developers need to handle large-scale datasets or require extensive context in their applications.
Therefore, the LLM vendors should develop and provide tools that help developers accurately calculate and manage their token usage. These tools should further offer cost optimization strategies~\cite{chen2023frugalgpt} tailored to different models and usage scenarios. Moreover, providing detailed guidelines on model selection, including trade-offs between cost and performance, will help developers make informed decisions that align with their budget constraints.

General challenges such as safety and privacy are paramount when using AI services. 
For LLM vendors, it is imperative to ensure compliance with global data protection regulations, such as GDPR (General Data Protection Regulation). This involves providing more transparent data processing and model training workflows to meet regulatory requirements. Additionally, LLMs may become targets of malicious attacks, such as prompt injection~\cite{liu2023prompt}, where the model output is manipulated to generate harmful or misleading content. Therefore, researching and implementing effective defense mechanisms is critical to safeguard their applications.
For LLM developers, especially in high-stakes domains (such as healthcare, finance, and software development), manual review of model outputs before their practical use is essential to mitigate risks caused by hallucinations~\cite{ahmad2023creating,kang2023deficiency,liu2024exploring,zhang2024llm}. Furthermore, after constructing a prompt, developers should utilize OpenAI’s Moderations API\footnote{\url{https://platform.openai.com/docs/api-reference/moderations}} to assess whether the prompt contains harmful content, ensuring the safety and integrity of the generated outputs.

\subsection{GPT Builder}

To meet specific user needs, the OpenAI organization initially introduced the ChatGPT plugin, allowing developers to design plugins with various functionalities based on user requirements. Recently, OpenAI launched GPTs with similar functionality. 
However, due to their later release, there are fewer posts related to GPTs. Although OpenAI has deprecated plugins and recommends using GPTs as a replacement for users' personalized needs, the OpenAI developer forum still lists plugins as an important subforum and combines plugins and GPTs into a single subforum. Moreover, plugin development and GPT development share many similarities in terms of development goals, technical foundations, and user usage. Therefore, we call both ChatGPT plugins and GPTs ``GPT Builder" and analyze the related challenges.
LLM developers commonly face challenges during the development and testing phases of GPT Builders. These include selecting suitable development environments and tools, addressing technical errors (such as server or parsing issues), and managing the complexity of integrating multiple GPTs for specific tasks. During testing, challenges often arise from validation errors, incorrect functionality, and debugging failures.
Our findings indicate that 12.1\% of the challenges fall into this category. The detailed subcategories are discussed as follows.

\begin{figure}[htbp] 
    \centering     \includegraphics[width=0.8\textwidth]{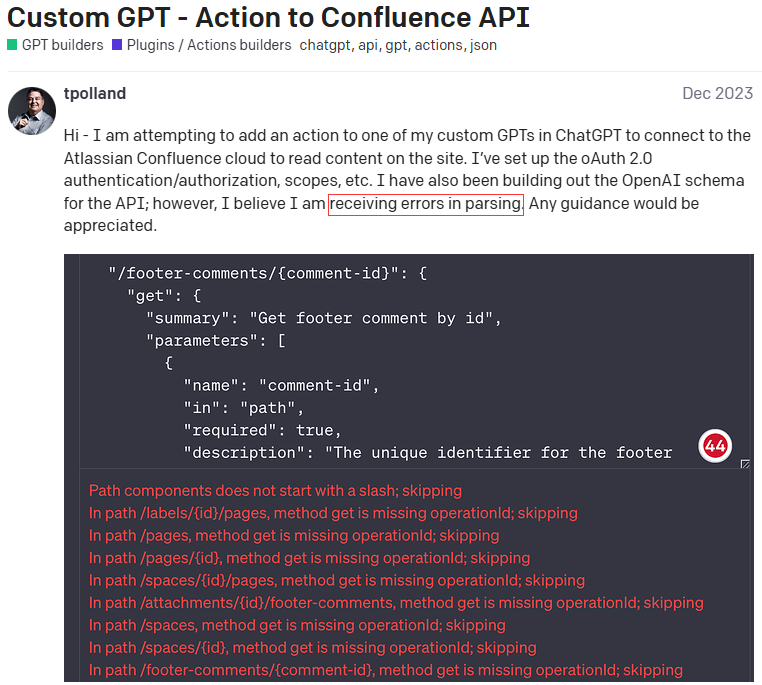} 
    \caption{A sample post in development (E.1).} 
    \label{Fig:post6} 
\end{figure}

\textbf{Development (E.1).}
In plugin development, developers encounter various challenges, including general and technical problems. 
Among general problems, developers seek guidance on selecting the most convenient IDE for plugin development\footnote{\url{https://community.openai.com/t/404743}} and whether Markdown can be integrated into plugin development\footnote{\url{https://community.openai.com/t/199399}}. Another category of challenges is related to technical issues. For instance, developers encounter a \emph{FatalServerError} while developing a plugin\footnote{\url{https://community.openai.com/t/297501}} or face parsing errors when attempting to connect GPTs to a cloud server\footnote{\url{https://community.openai.com/t/545528}}, as shown in Fig.~\ref{Fig:post6}. Additionally, developers ask how to use multiple GPTs simultaneously to handle the same task\footnote{\url{https://community.openai.com/t/552649}}. These challenges account for 11.2\% of the total challenges.

\textbf{Testing (E.2).}
After developing plugins or GPTs, developers conduct software testing to verify the correctness of the implemented functionalities. Throughout this process, they often encounter a range of challenges. For instance, the openapi.yaml file might fail initial validation because it makes an HTTPS request instead of HTTP when testing the plugin on a local server\footnote{\url{https://community.openai.com/t/254680}}, or developers encounter errors while testing if the plugin is working correctly\footnote{\url{https://community.openai.com/t/166239}}. In addition, some functional errors may be encountered during testing. For example, when developers test a GPT action, they find that after clicking to test, only the name of the GPT is displayed, without any other content output\footnote{\url{https://community.openai.com/t/597697/6}}. These challenges represent 0.9\% of the total challenges.

\finding{9}{
\textbf{Finding 9.} \emph{GPT Builder} accounts for 12.1\% of the total challenges. \emph{Development} dominates this category, accounting for 92.6\% of the challenges in \emph{GPT Builder}.}

\textbf{Discussion and Implication:} 
Developers face numerous general challenges when developing plugins and GPTs, particularly in selecting the appropriate development environments and tools. This involves decisions about which Integrated Development Environments (IDEs) to use, which programming languages best suit their project needs, and how to set up their development environments efficiently. Therefore, the OpenAI organization should provide comprehensive guides that cover the entire development lifecycle of plugins and GPTs, including recommendations for IDEs, programming languages, and other essential tools. This will help developers make informed choices that enhance their productivity and the quality of their developed products. Moreover, providing more presentative examples and templates can streamline the development process, particularly for those at varying skill levels. These resources can include step-by-step tutorials, sample projects, and best practices for developing and debugging, ensuring that developers have the support they need from project initiation to completion.

Following the development phase, rigorous testing is essential to ensure the proper functionality of plugins and GPTs.
Therefore, the OpenAI organization should provide comprehensive automated testing tools that allow developers to test their plugins and GPTs efficiently. Automated testing can help identify bugs early in the development cycle, reducing the time spent on manual testing.
Moreover, these testing tools should support multiple phases of the testing process, including unit testing, integration testing, and acceptance testing. By encompassing a comprehensive range of phases, developers can ensure their applications are both robust and reliable.

\subsection{Prompt}
Prompts play a pivotal role in LLM interactions, offering developers a mechanism to enhance the quality of the content produced. This category primarily focuses on challenges related to prompts, such as how to design and optimize them effectively. Developers frequently face issues in crafting prompts that achieve specific goals, modifying prompts to enhance output quality, and leveraging advanced techniques like retrieval-augmented generation, chain of thought, and in-context learning. Our findings indicate that 3.4\% of the challenges fall into this category. The detailed subcategories are discussed as follows.

\begin{figure}[htbp] 
    \centering     
    \includegraphics[width=0.9\textwidth]{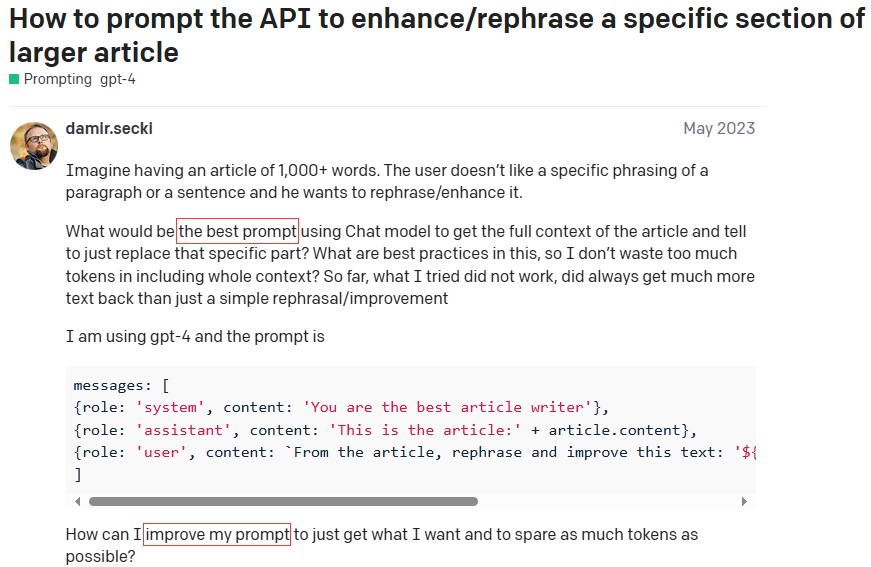} 
    \caption{A sample post in prompt design (F.1).} 
    \label{Fig:post7} 
\end{figure}

\textbf{Prompt Design (F.1).}
Developers often struggle with how to provide specific prompts to achieve the desired results and how to modify their existing prompts. This includes techniques like targeting specific topics and managing the length of generated content to guide the model more accurately. For example, they seek advice on prompts to regenerate part of the output\footnote{\url{https://community.openai.com/t/214339}}, which is shown in Fig.~\ref{Fig:post7}, and how to modify prompts to improve the conversational style with the GPT model\footnote{\url{https://community.openai.com/t/301950}}. Challenges related to prompt design constitute 2.3\% of the total challenges.

\textbf{Retrieval Augmented Generation (F.2).}
Retrieval Augmented Generation (RAG)~\cite{lewis2020retrieval} is a technique that enhances the quality and relevance of generated text by integrating information retrieval with generation processes. This technique combines the strengths of retrieval and generation, enabling the model to better grasp the context and produce text that meets user expectations. Developers encounter issues where RAG fails to improve output quality\footnote{\url{https://community.openai.com/t/550286}} or cause hallucinations after processing\footnote{\url{https://community.openai.com/t/408275}}. Questions also arise about preventing the RAG system from answering unrelated questions\footnote{\url{https://community.openai.com/t/434871}}. Challenges related to RAG account for 0.4\% of the total challenges.

\textbf{Chain of Thought (F.3).}
The Chain of Thought (CoT) technique~\cite{wei2022chain} is crucial for maintaining coherence and logical progression during text generation. It requires developers to craft input prompts that encourage the generation of text following a logical and semantic chain of thought. Achieving this involves careful selection of keywords, phrases, and sentence structures to steer the model toward producing coherent and structured text. Developers need a robust understanding of language, logical reasoning, and text generation capabilities, along with the skill to evaluate the model's behavior and outputs. Ensuring semantic coherence and meeting expectations might prompt questions on effectively utilizing the CoT technique to construct prompts\footnote{\url{https://community.openai.com/t/128180}} and seeking best practices for CoT in specific scenarios\footnote{\url{https://community.openai.com/t/17367}}. Challenges associated with the COT technique account for 0.2\% of the total challenges.

\begin{figure}[htbp] 
    \centering 
    \includegraphics[width=0.82\textwidth]{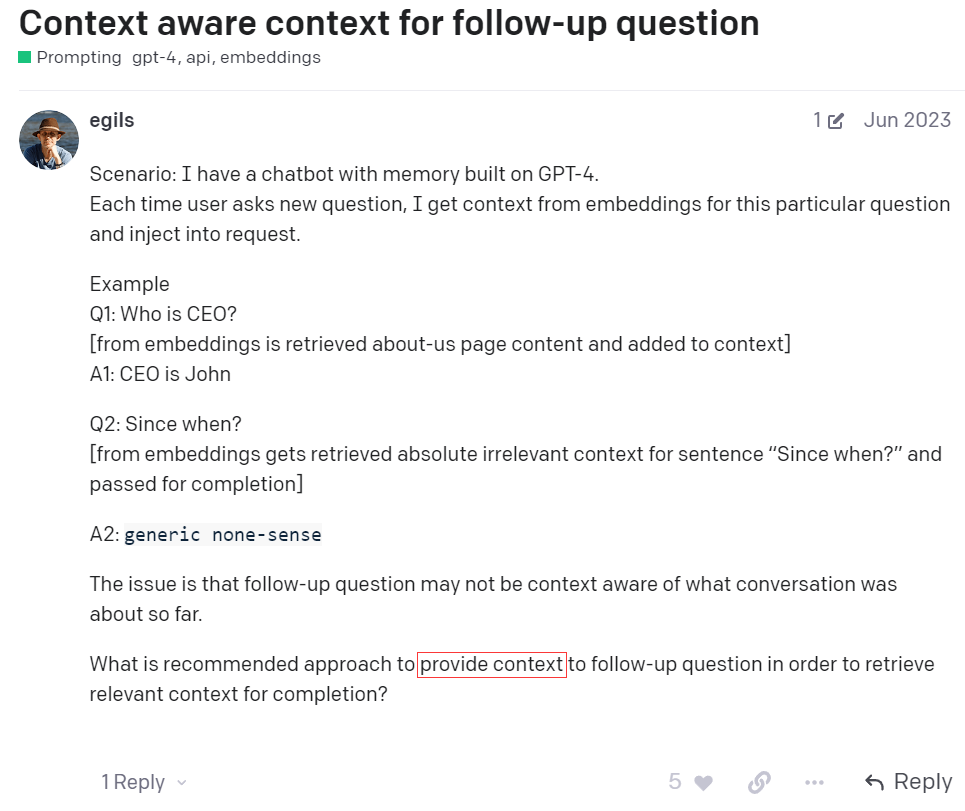} 
    \caption{A sample post in in-context learning (F.4).} 
    \label{Fig:post10} 
\end{figure}

\textbf{In-context Learning (F.4).}
In-context learning (ICL)~\cite{dong2022survey} involves the LLM's ability to learn and interpret based on the contextual information supplied within the input prompt. By meticulously designing these prompts, developers can embed specific contextual clues, enabling the model to deduce and produce content reflective of that context. For instance, there may be situations where, despite being provided with examples, the model repetitively generates identical outputs\footnote{\url{https://community.openai.com/t/17866}}. Developers seek advice on the recommended approach to provide context to follow-up questions to retrieve relevant context for completion\footnote{\url{https://community.openai.com/t/280666}}, as shown in Fig.~\ref{Fig:post10}. Challenges related to in-context learning represent 0.2\% of the total challenges.

\textbf{Zero-shot Prompting (F.5).}
Zero-shot prompting~\cite{kojima2022large} allows developers to leverage the model's capabilities without needing sample data. Common issues include comparing the efficacy of zero-shot prompting versus fine-tuning for specific tasks\footnote{\url{https://community.openai.com/t/289714}}. Recently, some developers observed that zero-shot prompting works as expected on the OpenAI Playground but fails to generate valid responses when called via the Python API\footnote{\url{https://community.openai.com/t/216108}}. 
Challenges related to zero-shot prompting constitute 0.2\% of the total challenges.

\textbf{Tree of Thoughts (F.6).}
Tree of Thoughts (ToT)~\cite{yao2024tree} is an advanced structured Chain of Thought (CoT) technique that guides the model toward generating text that is hierarchically organized and logically coherent. By creating detailed input prompts that integrate information from various branches and nodes within the ToT framework, developers enable the model to produce outputs with a clear hierarchy and logical structure. This technique assists the model in adhering to the ToT's structure, facilitating the creation of more organized and logically structured outputs. Researchers have found that ToT significantly enhances language models’ problem-solving abilities in tasks requiring non-trivial planning or search, such as the Game of 24, creative writing, and mini crosswords\footnote{\url{https://community.openai.com/t/226512}}. Challenges related to the tree of thoughts account for 0.1\% of the total challenges.

\finding{6}{
\textbf{Finding 6.} 3.4\% of the total challenges are related to \emph{Prompt}. \emph{Prompt Design} constitutes 67.6\% of this category, which is the highest proportion among the six subcategories.}

\textbf{Discussion and Implication:} 
In recent software engineering studies, advanced prompt design techniques like Chain-of-Thought and in-context learning~\cite{sun2024source,lu2023assessing,zhao2024automatic} have been shown to significantly improve the performance of software engineering tasks when using LLMs.
However, from our analysis of the OpenAI developer forum, we find that these advanced techniques have not been sufficiently utilized by developers (as evident from the relatively few questions concerning methods like Chain-of-Thought), indicating low usage. This finding suggests a gap in developer awareness or the perceived usability of these advanced prompt design techniques. We believe this observation has broader implications for the software engineering community beyond OpenAI.
For example, in addition to existing API documentation, the SE community should offer more detailed use cases with online courses, or technical tutorials, particularly for advanced prompt engineering techniques. Then developing tools that assist developers in automatically generating or optimizing prompts, or in evaluating different prompt strategies across tasks, could greatly enhance their ability to leverage advanced prompting techniques. Finally, the OpenAI forum should encourage developers to share best practices and lessons learned in using advanced prompting techniques.

\section{RQ4: Generalization of Our Methodology}
\label{sec:github}

\begin{figure}[htbp] 
    \centering
    \includegraphics[width=0.96\textwidth]{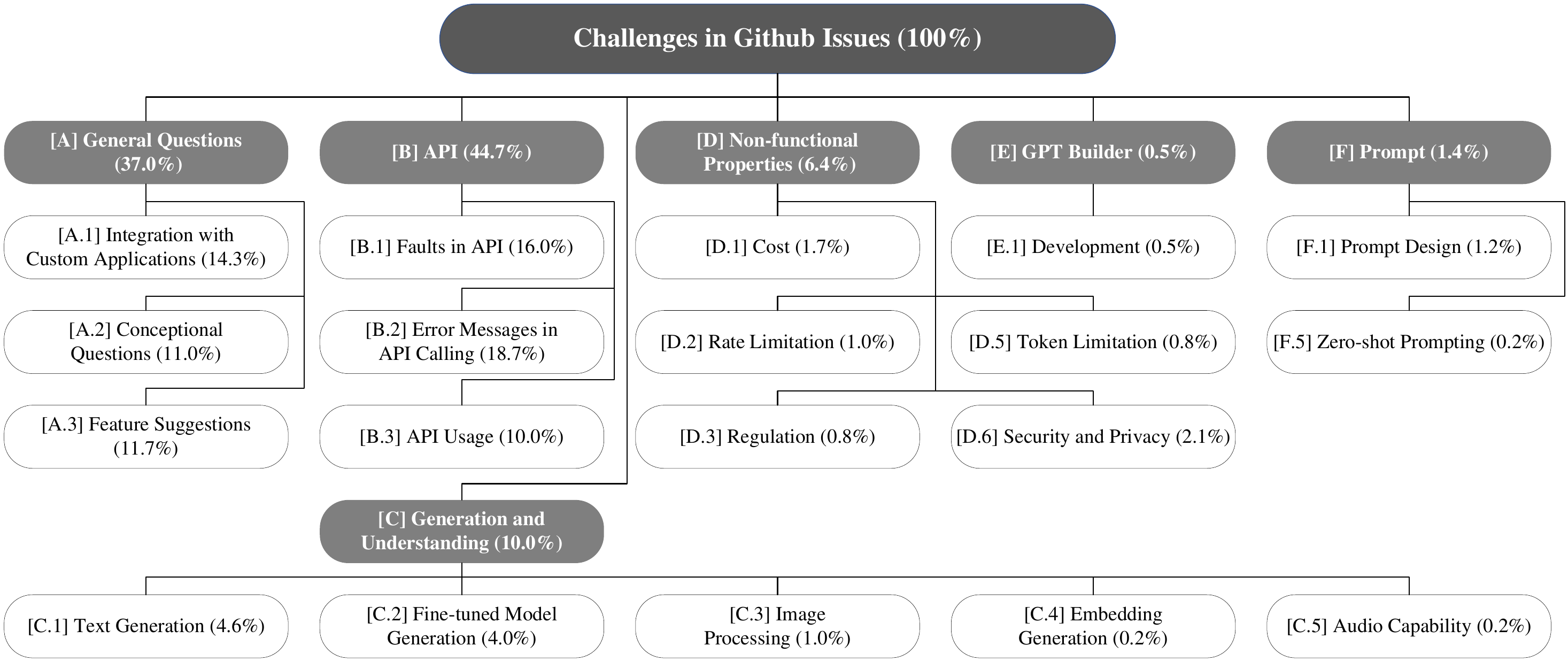} 
    \caption{Challenge Taxonomy for LLM developers Built from GitHub Issues.} 
    \label{fig:github} 
\end{figure}

To validate the generalization of our methodology, we further analyzed LLM-related issues on GitHub to construct the taxonomy. We show the process as follows.
First, we focus on the official GitHub accounts of three LLM vendors: Meta AI\footnote{\url{https://www.meta.ai}}, Google\footnote{\url{https://www.google.com}}, and OpenAI. The representative LLMs of these three vendors are Llama, Gemini, and GPT. We then select projects hosted on their homepages based on the following criteria: (1) relevance to LLMs, and (2) more than 50 issues.
After determining the projects, we download the issues from these specific repositories via GitMiner\footnote{\url{https://github.com/UnkL4b/GitMiner}}. Specifically, the data gathered for each issue includes its URL, title, submission date, status, labels, and issue description. 
In summary, we obtain 1,948 issues related to Llama, 560 issues related to Gemini, and 2,317 issues related to GPT.
Second, we randomly sample 356 posts from these 4,825 issues with a 95\% confidence level and $\pm$5\% confidence interval~\cite{aghajani2019software}. Then we manually analyze these sampled issues by using the open coding procedure shown in Section~\ref{sec:methodology}. 
Finally, we construct a corresponding taxonomy shown in Fig.~\ref{fig:github}.

In general, the taxonomy constructed based on the OpenAI forum covers the categories identified in the taxonomy built from GitHub. This suggests that the taxonomy developed from the OpenAI forum is representative to some extent. However, our analysis reveals some differences between the two taxonomies, which provide insights into the unique characteristics of these platforms and their user bases.

First, the [A.3] \emph{Feature Suggestions} category accounts for 11.7\% in the GitHub taxonomy, significantly higher than the 2.9\% observed in the OpenAI developer forum. This disparity highlights the collaborative nature of GitHub as a platform where technically proficient developers not only use existing features but actively contribute to the improvement and evolution of software and tools. These developers often engage in discussions about potential enhancements and new functionalities for tools, reflecting a proactive role in shaping the development landscape. In contrast, the OpenAI developer forum serves a broader audience, including beginners and normal users, where discussions are often centered around usage challenges rather than feature proposals.

Second, the [B] \emph{API} category constitutes 44.7\% of the GitHub taxonomy compared to 22.9\% in the OpenAI developer forum. This significant difference underscores the emphasis GitHub users place on API-related issues, including usage, optimization, and error handling. This focus is indicative of a user base that is deeply involved in software development and debugging processes, where APIs are integral to their daily tasks. The prevalence of API discussions on GitHub reflects the critical role APIs play in collaborative software projects and the need for highly technical solutions.

Third, issues related to [E] \emph{GPT Builder} are rarely discussed in the GitHub taxonomy, whereas they represent a notable category in the OpenAI developer forum. This discrepancy can be attributed to the fact that LLMs like Gemini and Llama do not support user plugins or GPT development in the same way as ChatGPT, leading to a lower frequency of GPT Builder-related discussions on GitHub.

Finally, some categories identified in the OpenAI forum taxonomy are absent from the GitHub taxonomy. For instance, [D.4] \emph{Promotion} is rarely discussed on GitHub, likely because the platform is centered around collaborative coding rather than promoting individual tools or plugins. Similarly, advanced prompt design techniques such as [F.2] \emph{Retrieval Augmented Generation} and [F.4] \emph{In-context Learning} are underrepresented, suggesting that these methods have not yet gained widespread attention among developers on GitHub. Finally, categories like [C.6] \emph{Vision Capability} may be limited due to the relatively small number of projects focused on such specific applications on GitHub.

These differences highlight the unique characteristics and developer priorities on each platform. GitHub caters to a more technical and collaborative audience, leading to a focus on API optimization and feature suggestions. In contrast, the OpenAI forum accommodates a diverse range of users, resulting in broader discussions that include usage challenges and advanced techniques. 

\section{Discussions}
\label{sec:threat}

In this section, we first review and synthesize the key findings, offer implications from these findings, and highlight opportunities for future work. We second analyze potential threats to our empirical study.

\subsection{Findings and Implications of Our Empirical Study}

This study analyzed the challenges faced by LLM application developers, constructing a taxonomy based on forum discussions to capture the breadth of these challenges. 
Our key findings can be summarized as follows:

\begin{itemize}
    \item \textbf{Diverse Developer Challenges.} The taxonomy reveals a wide range of challenges developers face, spanning functional and non-functional aspects of LLM development. Key categories include issues with API usage, GPT Builder development, prompt design, and advanced techniques.
    \item \textbf{Significance of API-Related Issues.} API-related challenges dominate the taxonomy, reflecting their central role in developer workflows. These include technical difficulties such as error handling, parameter configuration, and performance optimization, underscoring the need for robust support and documentation.
    \item \textbf{Emerging Interest in GPT Builders.} The presence of GPT Builder challenges highlights the growing interest in customizing LLM functionality. Developers frequently encounter issues during the development and testing phases, suggesting a need for more targeted resources in this area.
    \item \textbf{Underutilization of Advanced Prompt Techniques.} Despite their potential to enhance LLM outputs, advanced techniques like CoT and RAG remain underutilized, indicating gaps in developer awareness and accessibility.
    \item \textbf{Non-Functional Concerns Are Pervasive.} Challenges related to cost management, rate limitations, and data privacy reflect the importance of non-functional properties in LLM usage. These issues often act as barriers to adoption, especially for developers working with limited resources.
\end{itemize}

Based on these findings, we provide the following key actionable implications for LLM-related
stakeholders.

\begin{itemize}
    \item \textbf{Enhanced Documentation and Resources.} To address the diverse challenges identified, LLM vendors should prioritize the development of comprehensive documentation, including tutorials, best practices, and sample configurations. Specific focus should be given to API optimization, prompt engineering, and advanced techniques like CoT and RAG.
    \item \textbf{Tooling for Non-Functional Properties.} Non-functional concerns such as cost and rate limitations necessitate tools for real-time tracking and optimization. Developing related tools for monitoring usage and predicting costs could significantly improve developer experiences.
    \item \textbf{Supporting GPT Builder Development.} The prevalence of GPT Builder challenges underscores the need for streamlined development and testing tools. LLM vendors could introduce IDEs or automated testing frameworks tailored for GPT customization.
    \item \textbf{Promoting Advanced Techniques.} The limited use of advanced techniques suggests a need for targeted education initiatives. LLM vendors and the broader software engineering community should provide workshops, case studies, and automated prompt design tools to promote their adoption.
\end{itemize}

\subsection{Threats to Validity}

In this section, we discuss the threats to the validity of our empirical study, ensuring a comprehensive understanding of potential limitations.

\subsubsection{ Internal Validity}

\textbf{Expertise in constructing the challenge taxonomy:} The threat is that insufficient expertise among annotators constructing the taxonomy could lead to an inaccurate reflection of the challenges developers face. To alleviate this threat, we require that those involved in the taxonomy development have relevant experience in LLM development and a foundational understanding of large language models. This ensures the taxonomy’s reliability and relevance.

\textbf{Manual labeling of topics:} The threat is that manual labeling of topics could introduce subjective bias. To alleviate this threat, we employ a two-step labeling process: two independent annotators initially label the questions, and a third annotator, a domain expert, reviews these labels. Discrepancies are resolved through discussions among the annotators, thus reducing labeling bias to an acceptable level.

\subsubsection{External Validity}

\textbf{Limitations of OpenAI LLMs:} 
The threat is that focusing primarily on OpenAI’s LLMs in our empirical study may limit the generalization of our findings to other LLM vendors. 
First, by analyzing the number of followers on GitHub accounts for LLMs, we found that ChatGPT has 86.6k followers, while Llama and Gemini have only 5.2k and 1k followers, respectively. Therefore, it can be inferred that the LLMs provided by OpenAI are currently more popular.
Second, although different LLM vendors may offer varying features (such as image or audio processing capabilities), we believe that many of the challenges identified in our study (such as API usage, performance optimization, and security issues) are common across different LLMs. This implies that the insights from our study can inform development practices beyond OpenAI's ecosystem.
Finally, to demonstrate the generalizability of our constructed taxonomy, we further analyze challenges related to two other popular LLMs (such as Llama and Gemini) by mining issues from relevant projects on GitHub.

\textbf{Scope of data}: An external challenge arises from the temporal limitation of our data, encompassing posts up to June 2024. As the LLM vendors continue to update their products, new challenges will likely emerge, which our current empirical study will not capture.

\subsubsection{Construct Validity}

\textbf{Determining question difficulty levels:} 
The threat is that the method used to determine the difficulty level of questions posed on the OpenAI forum may not accurately capture the actual difficulty of the questions. To alleviate this threat, we introduce additional metrics beyond the proportion of posts with an accepted solution, such as the average time to receive the first reply and the distribution of posts by the number of replies. This analysis from multiple perspectives provides a more comprehensive view to measure question difficulty.

\textbf{Determining the popularity of categories:}
In our empirical study, we use the number of new posts as a proxy for measuring the popularity of categories. However, this metric may not fully reflect true popularity. For instance, older posts tend to accumulate more views over time, potentially introducing bias and making it challenging to accurately evaluate the popularity of newer posts or categories. An alternative approach could involve using the number of views as a metric to measure category popularity. However, the forum only provides cumulative view counts for each post, without the ability to differentiate views by year. As a result, this limitation prevents us from adopting the view-based metric for our analysis. 
\section{Related Work}
\label{sec:related}

\subsection{LLM for Software Engineering}
\label{sec:relatedLLM4SE}

With the recent advancements in LLMs, there has been growing interest in applying these models to software engineering research, a field now referred to as LLM4SE~\cite{hou2023large,fan2023large}. This area focuses on utilizing LLMs to enhance various software engineering activities. The primary activities that have garnered significant attention include software development, software quality assurance, and software maintenance. In this subsection, we summarize the recent research progress in LLM4SE based on these three core activities.

For the software development activity, the key tasks include code generation, code completion, and code summarization. 
Specifically,
code generation involves automatically generating source code from task descriptions. 
For example,  
Yan et al.~\cite{yan2023closer} explored ChatGPT's ability to solve programming competition problems under zero-shot learning.
Gu et al.~\cite{gu2024effectiveness} conducted an in-depth study of LLMs in domain-specific code generation, focusing on domains such as web, game, and math. To effectively incorporate domain knowledge, they utilized techniques like external knowledge inquirer, chain-of-thought prompting, and chain-of-thought fine-tuning.
Liu et al.~\cite{liu2024your,liu2024refining} examined and detailed various issues related to the quality of code generated by ChatGPT. 
Liu et al.~\cite{liu2024reliability} conducted a comprehensive evaluation of eight LLMs on five datasets to identify the challenges and limitations they face in different code generation scenarios.
Code completion is the task of automatically suggesting the next portion of code based on the current context.
For example,
Eghbali et al.~\cite{eghbali2024hallucinator} improved LLM-based code completion by retrieving suitable API references and iteratively querying the model with increasingly appropriate context information in the prompt. 
Tang et al.~\cite{tang2023domain} incorporated domain-specific knowledge into LLMs by leveraging external information sources for code completion, without requiring fine-tuning of the underlying models.
Li et al.~\cite{li2023cctest} used program structure-consistent mutations to test LLM-based code completion systems. These mutated inputs were then used to trigger potential errors in the code completion system.
Code summarization is the task of automatically generating concise descriptions or explanations of code snippets. 
For example,
Geng et al.~\cite{geng2024large} used LLMs to generate comments with multiple intents with in-context learning.
Ahmed et al.~\cite{ahmed2024automatic} improved LLM-based code summarization by augmenting the prompt with semantic information.
Sun et al.~\cite{sun2024source} conducted a comprehensive study on code summarization in the era of LLMs, examining various aspects such as prompt engineering techniques and temperature parameter settings to optimize performance.
Zhao et al.~\cite{zhao2024automatic} improved LLM-based smart contract comment generation achieved through in-context learning.

For the software quality assurance activity, the key tasks include vulnerability detection, test generation, and bug localization. Specifically,
vulnerability detection aims to identify security flaws or weaknesses in software code that could be exploited by malicious actors.
For example, 
Lu et al.~\cite{lu2023assessing} assessed the effectiveness of prompt tuning for LLM-based vulnerability detection.
Then they~\cite{lu2024grace} improved LLM-based vulnerability detection with graph structure and in-contex learning.
Steenhoek et al.~\cite{steenhoek2024dataflow} leveraged a data flow analysis-inspired graph learning framework to improve LLM-based vulnerability detection.
Chen et al.~\cite{chen2023chatgpt} investigated the strengths and limitations of utilizing ChatGPT for detecting smart contract vulnerabilities.
Test generation refers to the automatic generation of test cases or test scripts based on software requirements or existing code. 
For example, 
Schäfer et al.~\cite{schafer2023empirical} examined the effectiveness of LLMs in generating unit tests, where the prompts of LLMs include the function's signature and implementation, alongside usage examples extracted from the documentation.
Yuan et al.~\cite{yuan2023no} performed a quantitative analysis and user study to systematically evaluate the quality of tests generated by ChatGPT, focusing on aspects such as correctness, sufficiency, readability, and usability.
Ryan et al.~\cite{ryan2024code} proposed a code-aware prompting strategy for LLMs in test generation.
Bug/fault localization refers to the process of automatically identifying the specific location or cause of a bug in the code. 
For example, 
Wu et al.~\cite{wu2023large} provided an in-depth investigation into ChatGPT's capabilities for fault localization, considering the effects of prompt engineering and code context length.
Kang et al.~\cite{kang2024quantitative} proposed an LLM-based fault localization technique, which can generate both an explanation of the bug and a suggested fault location. 
Hossain et al.~\cite{hossain2024deep} used LLMs to identify bug locations at the token level and then applied them to perform bug fixing.

For the software maintenance activity, the key tasks include program repair, code clone detection, and code review. Specifically,
program repair refers to the automated process of identifying and fixing bugs/vulnerabilities in source code. For example, 
Jin et al.~\cite{jin2023inferfix} enhanced LLM-based program repair by incorporating a static analyzer. Their approach involves a transformer encoder model, pretrained using contrastive learning, which is specifically designed to identify semantically similar bugs and corresponding fixes. Additionally, a decoder model is fine-tuned on supervised bug-fix data to generate effective repairs.
Peng et al.~\cite{peng2024domain} proposed a prompt-based approach for repairing Python type errors. This method begins by mining generalized fix templates, which serve as domain knowledge. These templates are then used to generate code prompts, guiding the model to produce accurate repairs for type errors.
Code clone detection is the task of identifying identical or highly similar code fragments within or across software systems. 
For example, 
Dou et al.~\cite{dou2023towards} conducted the first comprehensive evaluation of LLMs for code clone detection, examining various clone types, programming languages, and prompt strategies.
Zhang et al.~\cite{zhang2024assessing} assessed the performance of ChatGPT for code clone detection. For example, they found that LLMs exhibited low effectiveness in detecting the most complex Type-4 code clones.
Code review is a critical step in software quality assurance, involving tasks such as reviewer recommendation, code issue prediction, and review comment recommendation and generation. 
Li et al.~\cite{li2022automating} proposed a pre-trained model designed specifically for code review tasks, such as code change quality estimation, review comment generation, and code refinement.
Yu et al.~\cite{yu2024fine} aimed to improve the accuracy and comprehensibility of code review by fine-tuning LLMs.

\subsection{Studies on ML/DL Developers' Challenges}

In traditional software development, developers face a broad range of challenges across different application domains such as mobile applications~\cite{rosen2016mobile}, web applications~\cite{scoccia2021challenges}, and concurrency programs~\cite{ahmed2018concurrency}. These challenges include understanding complex requirements, ensuring code quality and maintainability, handling bugs/security vulnerabilities, and integrating various software components.

The rapid development of machine learning (ML) technologies poses new challenges for software developers. Thung et al.~\cite{thung2012empirical} analyzed bug severity, bug fixing efforts, and bug impacts in ML systems. Alshangiti et al.~\cite{alshangiti2019developing} demonstrated that questions in the process of ML application development are more difficult to answer than questions in other domains on Stack Overflow. 
These studies highlight the unique difficulties in developing and maintaining ML systems, such as data quality issues, model training complexities, and integrating ML models with traditional software components.

In the realm of deep learning (DL), DL developers face distinct challenges. For example, Zhang et al.\cite{zhang2019empirical} found that program crashes, model migration, and implementation questions are the top three most frequently asked questions when developing DL applications. 
Morovati et al.~\cite{morovati2024common} investigated the challenges of deep reinforcement learning application development.
Other researchers have characterized faults in software that make use of DL frameworks. For example, Zhang et al.\cite{zhang2018empirical} categorized the symptoms and root causes of these DL bugs for TensorFlow and proposed strategies to detect and locate them. Then Islam et al.\cite{islam2019comprehensive} and Humbatova et al.\cite{humbatova2020taxonomy} extended the scope to include bugs in programs written based on more popular DL frameworks (such as Caffe, Keras, TensorFlow, Theano, and PyTorch), presenting more comprehensive results. Recently, Shen et al. also analyzed the DL compiler bugs~\cite{shen2021comprehensive,chen2023toward}.

Empirical studies also highlighted specific issues, such as bugs in model optimization~\cite{guan2023comprehensive}, compatibility problems~\cite{wang2023compatibility,guo2019empirical}, DL deployment challenges~\cite{chen2020comprehensive}, and performance-related concerns~\cite{cao2022understanding}.

In recent years, the rise of foundation models, particularly large language models, has led to remarkable progress across numerous domains, including software engineering. However, developers working with LLMs face a distinct set of challenges. Different from conventional ML/DL development, which often focuses on model training and optimization, LLM development introduces new issues, such as prompt design, the hallucination problem, result reproducibility, and API call costs. These challenges, combined with the multi-modal nature of LLMs (e.g., handling text, images, and audio), create complexities that are not typically encountered in traditional ML/DL workflows. Unfortunately, due to the rapid pace of development in this field, there are currently no comprehensive studies addressing these emerging challenges or offering clear guidelines for LLM developers. This study aims to fill this gap by conducting the first large-scale empirical study into the specific challenges faced by LLM application developers. By manually analyzing {\samplednum} questions from the OpenAI developer forum, this study constructs a taxonomy of LLM-related challenges and provides actionable insights for developing LLM-driven systems.

\section{Conclusion}
\label{sec:conclusion}

In this study, we conduct a comprehensive analysis of the challenges faced by LLM developers. By examining relevant posts on the OpenAI developer forum, we observe that LLM-related development is gaining significant traction, with developers encountering numerous challenging issues compared to traditional software development. Our goal is to analyze the underlying challenges reflected in these questions. We investigate the popularity trends on the LLM developer forum and the difficulty levels of the problems raised by developers. Subsequently, we manually inspect {\samplednum} sampled posts related to LLM development and construct a detailed taxonomy consisting of 6 main categories and 26 subcategories, representing the challenges that LLM developers encounter. Finally, we discuss the implications of our findings for various stakeholders, including LLM developers and LLM vendors.

For future research, there are several avenues to explore. First, expanding the analysis to include posts from other platforms, such as Stack Overflow, can provide a more comprehensive view of the challenges faced by LLM developers. Second, an in-depth study of the solutions proposed in response to the identified challenges can yield valuable insights into effective strategies and best practices. Finally, automating the categorization process using machine learning techniques can improve the efficiency and accuracy of the analysis.

\begin{acks}
The authors would like to thank the editor and three anonymous reviewers for their insightful comments and suggestions, which can substantially improve the quality of this work.
Xiang Chen and Chaoyang Gao have contributed equally
to this work and they are co-first authors.
Xiang Chen and Yong Liu are the corresponding authors. 
This research was partially supported by 
the National Natural Science Foundation of China (Grant no. 61202006) and the Postgraduate Research \& Practice Innovation Program of Jiangsu Province (Grant no. SJCX24\_2022).
\end{acks}

\bibliographystyle{ACM-Reference-Format}
\bibliography{sample-base}

\appendix

\end{document}